\documentclass[fleqn,usenatbib]{mnras}

\usepackage{newtxtext,newtxmath}

\usepackage[T1]{fontenc}

\DeclareRobustCommand{\VAN}[3]{#2}
\let\VANthebibliography\thebibliography
\def\thebibliography{\DeclareRobustCommand{\VAN}[3]{##3}\VANthebibliography}

\usepackage{graphicx}	
\usepackage{amsmath}	
\usepackage{subcaption}
\usepackage{orcidlink}

\def\gtorder{\mathrel{\raise.3ex\hbox{$>$}\mkern-14mu
             \lower0.6ex\hbox{$\sim$}}}
\def\ltorder{\mathrel{\raise.3ex\hbox{$<$}\mkern-14mu
             \lower0.6ex\hbox{$\sim$}}}

\title[Core-Collapse Analysis]{The Late Time Optical Evolution of Twelve Core-Collapse Supernovae: Detection of Normal Stellar Winds}

\author[M. Rizzo Smith et al.]{
M.  Rizzo Smith\orcidlink{0000-0001-6721-4909}$^{1}$\thanks{E-mail: rizzosmith.1@osu.edu},
C. S. Kochanek$^{1,2}$,
J.~M.~M.~Neustadt$^{1}$
\\
$^{1}$Department of Astronomy, The Ohio State University, 140 West 18th Avenue, Columbus OH 43210\\
$^{2}$Center for Cosmology and AstroParticle Physics, The Ohio State University, 191 W. Woodruff Avenue, Columbus OH 43210\\
}

\date{Accepted XXX. Received YYY; in original form ZZZ}

\pubyear{2022}

\begin{document}
\label{firstpage}
\pagerange{\pageref{firstpage}--\pageref{lastpage}}
\maketitle

\begin{abstract}
We analyze the late time evolution of 12 supernovae (SNe) occurring over the last ${\sim}$41 years, including nine Type IIP/L, two IIb, and one Ib/c, using UBVR optical data from the Large Binocular Telescope (LBT) and difference imaging. We see late time (5 to 42 years) emission from nine of the eleven Type~II SNe (eight Type IIP/L, one IIb). We consider radioactive decay, circumstellar medium (CSM) interactions, pulsar/engine driven emission, dust echoes, and shock perturbed binary companions as possible sources of emission. The observed emission is most naturally explained as CSM interactions with the normal stellar winds of red supergiants with mass loss rates in the range $-7.9\lesssim \log_{10}(M_\odot\text{yr}{}^{-1}) \lesssim-4.8$. We also place constraints on the presence of any shock heated binary companion to the Type Ib/c SN~2012fh and provide progenitor photometry for the Type~IIb SN~2011dh, the only one of the six SNe with pre-explosion LBT observations where the SN has faded sufficiently to allow the measurement.  The results are consistent with measurements from pre-explosion Hubble Space Telescope (HST) images.

\end{abstract}



\section{Introduction}

For all massive stars ($>8M_\odot$), the end stages of stellar evolution involve the ultimate collapse of the iron core. In most cases this leads to an explosion and a luminous supernova (SN).  These core-collapse supernovae (ccSNe) are categorized into sub-types based on their photometric and spectroscopic properties. To simplify the typing somewhat, Type IIP/L SN arise from red supergiants (RSG), Type Ib/c SN come from stars stripped of their envelopes, and Type IIn SN explode into a dense circumstellar medium (CSM).  In the standard picture of non-Type~IIn explosions, 
SNe can support their near peak luminosities until the ejecta recombine and become optically thin, and then the luminosity in the nebular phase is from radioactive decay.  In Type~IIn SNe, there is a significant or dominant contribution to the emission from
shock heating the dense CSM.

As part of a search for failed SNe first proposed by \citet{Kochanek2008}, we have been monitoring 27 nearby ($<10$Mpc) galaxies to search for failed supernovae. The survey also enables two unique studies of the stars which do explode. First, it can be used to study the pre-SN variability of the progenitor stars, so far with null results almost down to the levels of the variability of local RSGs, (e.g., \citealt{2012ApJ...747...23S}, \citealt{2017MNRAS.467.3347K}, \citealt{Johnson2018}). This is difficult to reconcile with claims that SN light curves require unusually high pre-SN mass loss rates (e.g., \cite{2018ApJ...858...15M} and \cite{2022ApJ...930..119W} discuss many claimed examples). Second, by waiting for the SN to fade, it is possible to obtain high  precision UBVR photometry of the progenitor stars using image subtraction (e.g., \citealt{2017MNRAS.472.3115J}). Based on either SN 1987A (e.g., \citealt{2014ApJ...792...10S}) or typical estimates of the radioactive yields of ccSNe, this should be feasible after ${\sim}3$-$4$ years.

\begin{figure*}
    \centering
    \includegraphics[width=0.8\textwidth]{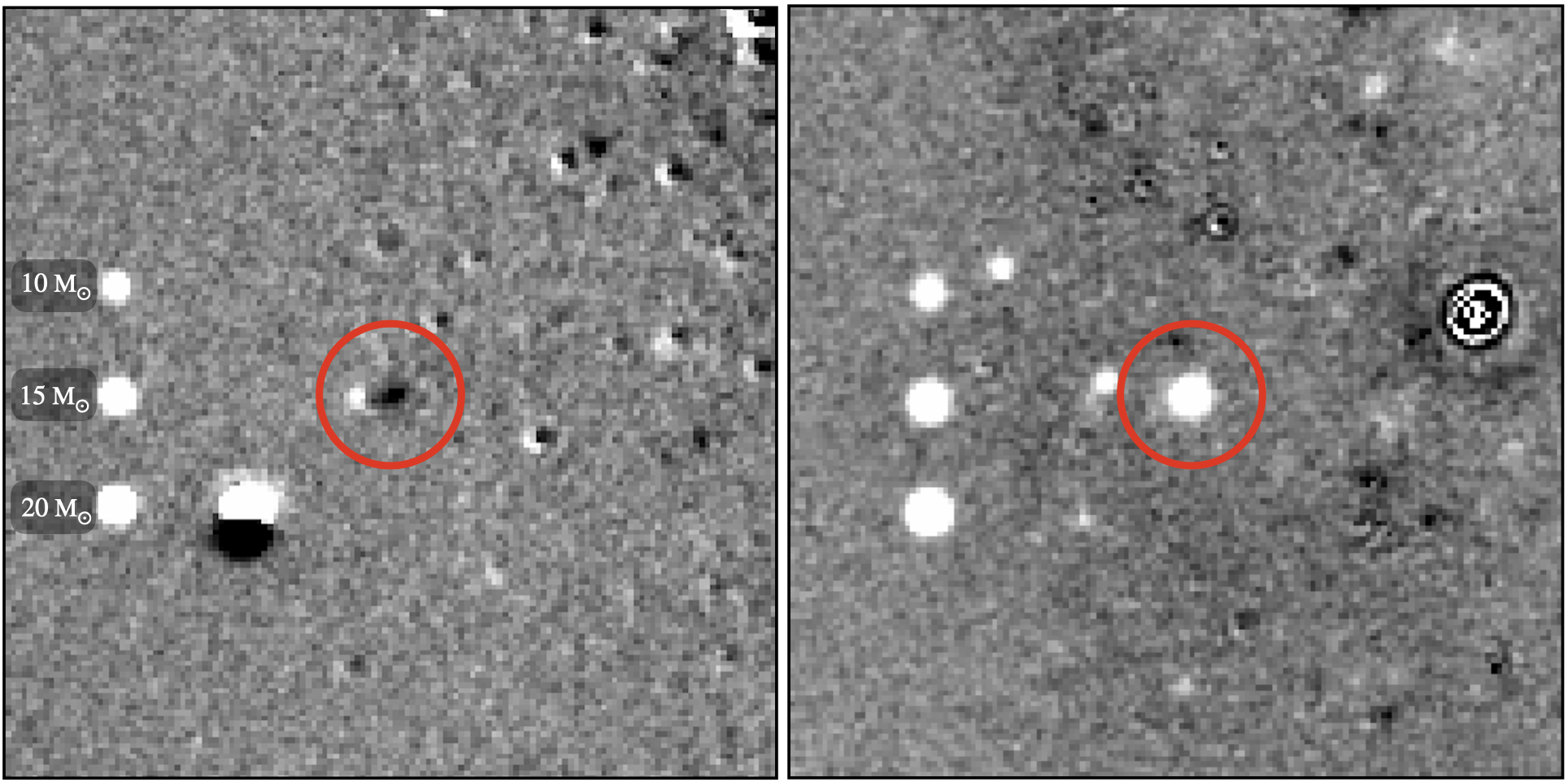}
    \caption{The average of the three most recent subtracted, LBT R band images of SN~2013ej (left) and
     SN~2011dh (right) with the SN position indicated by the 3\farcs0 circle. The three sources arranged
     vertically to the left of each SN are the expected signals for 10 (top), 15 (middle), and 20 $M_\odot$ (bottom) progenitors. While the subtractions for SN~2013ej are not the cleanest, it is presently more luminous than its progenitor and fading. }
\label{fig:13model}
\end{figure*}

In the most recent search for new failed SN, \citet{2021MNRAS.508..516N} noted that SN~2013am and SN~2013ej were still  brighter than their progenitors a decade after explosion. This is illustrated in Fig.~\ref{fig:13model}
for SN~2013ej and SN~20011dh, the one example showing the expected fading behaviour.  Here we are showing the difference
between the most recent images and pre-SN images, so a ``negative'' image of the progenitor appears once the SN becomes fainter than the progenitor.  This has clearly happened for SN~2011dh and the observed signal matches the flux expected for the progenitor from pre-SN HST images (see below, \citealt{2011ApJ...741L..28V},  \citealt{2011ApJ...739L..37M}).  Such a source clearly is not present for SN~2013ej,
and this is difficult to reconcile with radioactive decay as an energy source. Other possible sources of late time emission are interaction with circumstellar material (CSM, e.g., \citealt{1982ApJ...258..790C}), dust echoes (e.g., \citealt{1986ApJ...308..225C}), pulsar/engine driven emission (e.g., \citealt{2010ApJ...717..245K}), and shock heated companion stars (e.g., \citealt{2021MNRAS.505.2485O}).  

The puzzle of SN~2013am and SN~2013ej motivates this systematic examination of the late time emission from the 13 SNe listed in Table~\ref{tab:SN_INFO} and discussed in Appendix~\ref{appendix} that have occurred in the LBT  galaxies since 1980. Of these, 7 exploded during the course of the survey and so we should be able to observe the vanishing of the progenitor. For the other 6, we can only search for continuing emission. SN~2009dh is one of the $7$ in the LBT sample, but it is heavily obscured and suffers from poor
image subtractions.  We include it for completeness but do not consider it part of the main study.
Most of these SN are well studied. Seven have existing reports of CSM interactions and pre-SN mass loss constraints usually based on radio or X-ray observations, and five are reported to have either optical or infrared
dust echoes. These earlier results are summarized in Appendix~\ref{appendix}.  The new finding here is that late time (decade(s)) optical emission comparable to the luminosity of the progenitor stars appears to be the norm rather than the exception.
In \S~\ref{sec:observation} we describe the data, analysis methods, and the five potential sources of late time emission. We analyze the SNe in \S~\ref{sec:Discussion} and discuss the results in \S~\ref{sec:results}.

\begin{table*}
    \centering
    \begin{tabular}{lllrcccccc}
        \hline
        \multicolumn{1}{c}{SN} & \multicolumn{1}{c}{Type} & \multicolumn{1}{c}{Host} & \multicolumn{1}{c}{Distance} &   Distance & Galactic & Host & Host Extinction & Apparent R & Peak\\
            &   &  & \multicolumn{1}{c}{(Mpc)} & Reference &   $E(B-V)$    &   $E(B-V)_{host}$  &  Reference & Peak Mag & Reference\\
        \hline
        \hline
        SN 1980K    &   IIL    & NGC 6946  &  5.96 & 1  & 0.291 &  0.07  &    8 & 11.45 & 21 \\
        SN 1993J    &   IIb     & M81       & 3.65 & 2  & 0.069 &  0.13  &    9 & $10.47$ & 22 \\
        SN 2002hh   &   II-P    &   NGC 6946    &  5.96 & 1  &  0.290  &   1.63   &    10 & $15.54$ & 23 \\
        SN 2003gd   &   II-P    &   NGC 628     & 8.59 & 4  &   0.060  &  $0.07\pm0.06$   &    11  & $13.63$ & 24 \\
        SN 2004dj   &   II-P    &   NGC 2403    & 3.56 &  5 &   0.035  &   $0.026\pm0.002$   &   12 & $11.51$ & 25  \\
        SN 2005cs   &   II-P    &   NGC 5194    &  8.30 &  6 &  0.031  &   $0.01$   &    13  & 14.36 & 26\\
        \hline
        SN 2009hd   &   IIL    &   NGC 3627    &  10.62 &  3 &  0.029  &   $1.19\pm0.05$   &    14  & 15.75 & 27 \\
        SN 2011dh   &   IIb     &   NGC 5194    & 8.30 &  6 &   0.031  &   $0.04$   &    15  & 12.25 & 28  \\
        SN 2012fh   &   Ib/c    &   NGC 3344    & 6.90 &  7 &   0.028  &   $\simeq0$   &    16 & 16.24 & 29  \\
        SN 2013am   &   II-P    &   NGC 3623    & 10.62  &  3 &  0.021  &   $0.55\pm0.19$ & 17 & 15.59 & 30   \\
        SN 2013ej   &   IIL    &   NGC 628     &  8.59 & 4  &  0.060  &   $\simeq0$   &    18 & 12.43 & 31  \\
        SN 2016cok  &   II-P    &   NGC 3627    & 10.62 & 3  &   0.029  &   $0.50\pm0.02$   &    19 & 15.25 & 32  \\
        SN  2017eaw &   II-P    &   NGC 6946    & 5.96  &  1 &  0.297  &   $0.11\pm0.05$   &    20 & 12.44 & 31  \\
        \hline
    \end{tabular}
    \caption{The Galactic extinction from \citet{2011ApJ...737..103S}, for an assumed foreground reddening law of $R_V=3.1$.  The horizontal line separates SNe without pre-SN LBT data (top) from those with pre-explosion LBT data (bottom). The apparent peak magnitude for SN~1980K is the V band magnitude used for the dust echo analysis. References: (1) \citet{2000A&A...362..544K}, (2) \citet{2011ApJ...743..176G}, (3) \citet{N3627Dist}, (4) \citet{2008ApJ...683..630H}, (5) \citet{1997ApJS..109..333W}, (6) \citet{2009ApJ...694.1067P}, (7) \citet{2000A&A...356..827V}, (8) \citet{2012ApJ...749..170S}, (9) \citet{2022MNRAS.509.3235Z}, (10) \citet{2002IAUC.8024....1M}, (11) \citet{2005MNRAS.359..906H}, (12) \citet{2004IAUC.8384....3G}, (13) \citet{2007ApJ...662.1148B}, (14) \citet{2011ApJ...742....6E}, (15) \citet{2014A&A...562A..17E}, (16) \citet{2012ATel.4544....1M}, (17) \citet{2014ApJ...797....5Z}, (19) \citet{2014MNRAS.438L.101V}, (19) \citet{2017MNRAS.467.3347K}, (20) \citet{2019MNRAS.487..832B}, (21) \citet{Barbon1980}, (22) \citet{1994AJ....107.1022R}, (23) \citet{2007PZ.....27....5T}, (24) \citet{2005MNRAS.359..906H}, (25) \citet{2008PZ.....28....8T}, (26) \citet{2006MNRAS.370.1752P}, (27) \citet{2011ApJ...742....6E}, (28) \citet{2014A&A...562A..17E}, (29) \citet{Zheng2022}, (30) \citet{2014ApJ...797....5Z}, (31) \citet{2015ApJ...806..160B}, (32) \citet{2019MNRAS.490.2799D}, (33) \citet{2019MNRAS.487..832B} .}
    \label{tab:SN_INFO}
\end{table*}

\section{Observations and Models} \label{sec:observation}

The data are obtained using the Large Binocular Camera (LBC, \citet{Gig2008}) on the LBT \citep{HillLBT} in the U, B, V, and R bands with image subtraction done using \verb"ISIS" \citep{ISIS1999, ISIS2000}. The details of the data processing are given in \citet{Gerke2015}, \citet{2017MNRAS.469.1445A}, and \citet{2021MNRAS.508..516N}. For SNe with observations obtained pre-event, we built the reference images using the best observations before the explosion. With these reference images, difference imaging photometry will yield a ``negative" image of the progenitor as the SN fades completely. 

Lower quality data points are flagged  as those with seeing FWHM $> 1\farcs5$ or an ISIS flux scaling factor $< 0.8$. This scaling factor is the counts per unit flux compared to the reference image and thus is an estimate of the transparency for each epoch relative to a flux reference epoch. For the calibrations, we follow the same methods as \citet{Gerke2015}, and \citet{2017MNRAS.469.1445A}. Stars in the reference images are matched to the Sloan Digital Sky Survey (SDSS, \citealt{2012ApJS..203...21A}) and the corresponding \textit{ugriz} AB magnitudes are transformed into \textit{UBVR} Vega magnitudes following \citet{2006A&A...460..339J}. Following \citet{Johnson2018}, we extract light curves for both the target and a grid of comparison points to characterize the local noise. An inner grid of four points is placed 7 pixels apart (${\sim}$1\farcs6 given the plate scale of 0\farcs23 pixel$^{-1}$). The outer grid is placed 15 pixels away or ${\sim}$3\farcs5. Figures~\ref{fig:MultiStackIIP/L}-\ref{fig:MultiStackIb/c} show the resulting difference image light curves grouped by SN type. The formal errors are shown but are generally small which is why we look at the comparison points. We compute  band luminosities ($\nu L_\nu$) using the distances and extinctions in Table~\ref{tab:SN_INFO}. The extinction corrections for SN~2002hh are enormous, nearly 10 magnitudes in the U band.

We fit the late time (meaning well into the nebular phase) light curves as a linear function in luminosity ($\nu L_{\nu}$ for each band),
\begin{equation}
L =L_{SN} + \beta_{SN} (t-t_0)
\end{equation}
with $t_0$ being the mean age of the points being fit so
that there are no significant covariances between the two parameters $ L_{SN}$ and $\beta_{SN}$.  We do the same for the light curves
of the comparison points. Tables \ref{tab:SN Decay} and \ref{tab:SN Decay Cont.} report $ L_{SN}$ and $\beta_{SN}$ and their uncertainties
for each of the SNe.  From the comparison points we report the mean values $\langle L_{i} \rangle $ and $\langle \beta_{i} \rangle$ and the
dispersion of the values $\sigma_L$ and $\sigma_\beta$. Cases like SN~2009hd where there is 
significant dispersion in the comparison sample are a strong indication that the values are dominated by systematics.  As mentioned earlier, the image subtractions for SN~2009hd are
not very clean, driving the measured values and the large scatter.
We do not consider SN~2009hd further. 

The image subtraction light curves do not include the flux of the source in the reference image. For the SN with pre-SN images, this corresponds to the flux of the progenitor star.  For the other SN, it is 
simply the mean flux of the SN over the images used to construct the reference image.  We used aperture 
photometry to estimate the flux in the reference image, and the resulting luminosity is also reported in
Tables~\ref{tab:SN Decay} and~\ref{tab:SN Decay Cont.}.  Particularly in the case of the progenitor stars, crowding generally makes the flux unmeasurable, so the
luminosity estimates are dominated by systematic uncertainties.  This is the reason why we focus on the subtracted
light curves, which are little affected by crowding, separate from the reference flux.

Table~\ref{tab:HST_DATA} summarizes the cases where there are HST observations that can be used to explore this
normalization.  These are either pre-explosion observations for the SN with pre-SN LBT observations, or late time
observations that (nearly) overlap the LBT observations.
We converted the HST magnitudes to band luminosities using the HST zero points and the distances and extinctions from Table~\ref{tab:SN_INFO}, and these band luminosities are reported in Table~\ref{tab:HST_DATA}. We are interested in basic normalizations, not modest filter differences, so we simply view the HST filter as representing the closest equivalent LBT
filter (so, for example, V=F555W=F606W).  While we are frequently worried about line emission, we do not think these details represent a serious concern given the wavelength ranges of the filters. 

To characterize the late time photometric evolution of each of the SNe, we must
consider the expected properties of the progenitor star and then the 5 possible sources of late time luminosity:
radioactive decay, CSM interactions, dust echoes, neutron star spin down, and shock heated companions.

\subsection{Progenitor Stars}

While we consider a broader range of supernovae, the most interesting cases are the 6 where we can construct
a reference image from pre-supernova data.  For a reference image built from pre-SN images, we will be left
with an ``inverse image" of the progenitor once the SN has completely faded. SN~2011dh, shown in Fig.~\ref{fig:13model}, is 
the one case out of these 6 where this appears to have occurred.  We model its
spectral energy distribution (SED) following the procedures of \citet{2015MNRAS.452.2195A} using DUSTY \citep{1997MNRAS.287..799I, 1999astro.ph.10475I, 2001MNRAS.327..403E} so that we can consider self-obscuration and Solar metallicity stellar atmosphere models \citep{2003IAUS..210P.A20C}.

\begin{figure}
     \centering
   \includegraphics[width=0.45\textwidth]{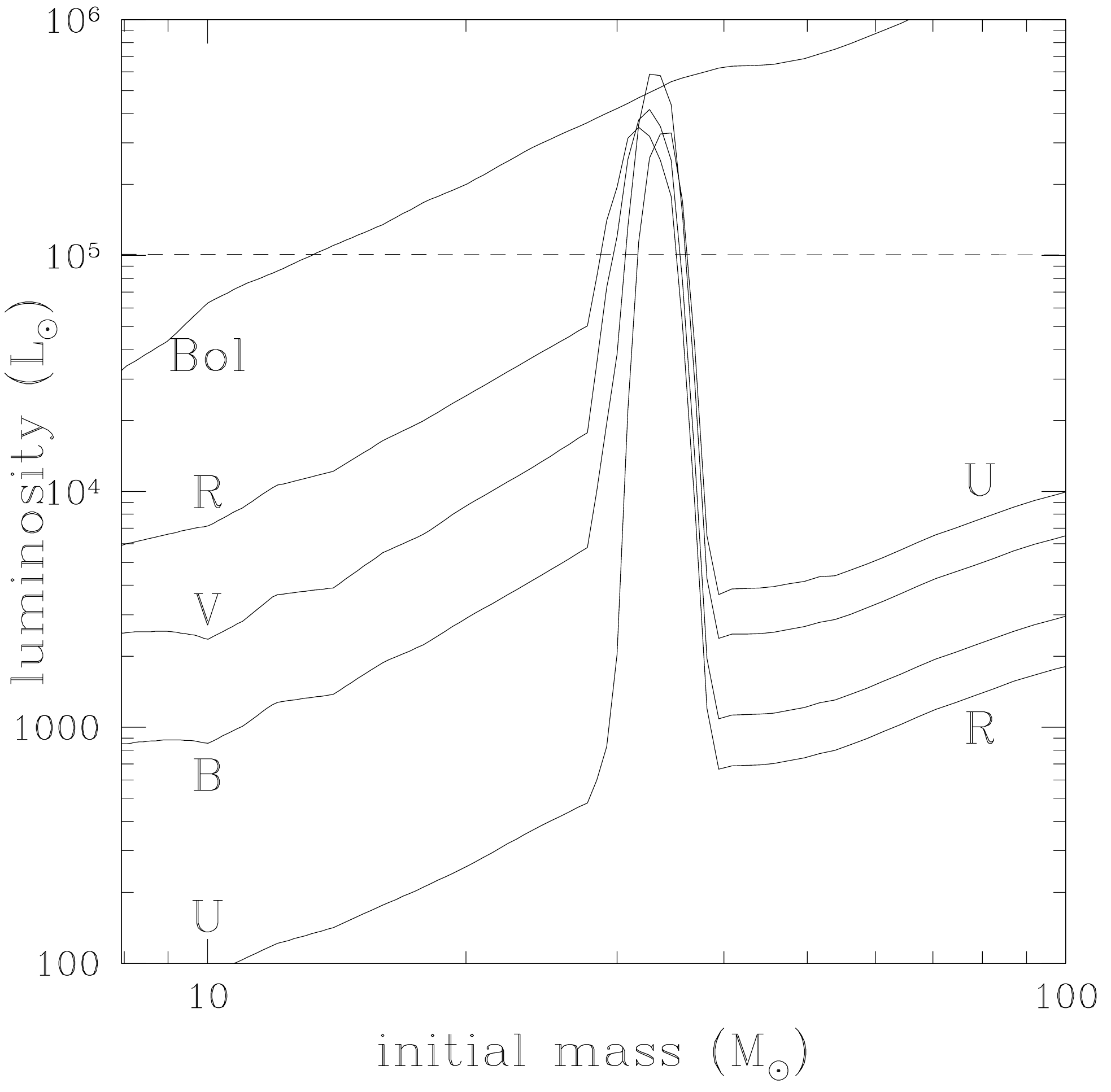}
   \caption{UBVR and bolometric luminosities of SN progenitors at death for the Solar metallicity PARSEC isochrones. The
 horizontal dashed line is the estimated luminosity of the progenitor of SN~2011dh.}     \label{fig:prog}
\end{figure}

  Figure~\ref{fig:prog} shows the expected UBVR band luminosities
along with the bolometric luminosity 
as a function of the initial progenitor mass $M_{ZAMS}$ for the Solar metallicity PARSEC \citep{2012MNRAS.427..127B, 2013MNRAS.434..488M} isochrones.  
The \citet{2013A&A...558A.131G} progenitor luminosities are similar.  The bolometric luminosity is roughly a power law in mass, with
\begin{equation}
\log L_{prog} \simeq 4.8 + 1.5 \log (M/10 M_\odot).
\label{eqn:lprog}
\end{equation}
In the PARSEC models, the stars are RSGs for $M_{ZAMS} \lesssim 30M_\odot$ so
the stars are faintest in the U band and brightest in the R band.  The mass loss monotonically increases with mass, so
there is a small transition region where the temperatures lead to the emission peaking in the optical, and then the
stars are very hot, heavily stripped stars that are brightest in the U band and faintest in the R band.  With binary
interactions, the temperatures at a given mass can be very different, but the bolometric luminosity is still basically set by the
helium core mass.

\subsection{Radioactive Decay}\label{sec:radioactive}

All SN will show emission due to radioactive decay, initially dominated by $^{56}$Ni, then $^{57}$Co 
and finally $^{44}$Ti (e.g., \citealt{2014ApJ...792...10S}).  A typical Type~IIP SN synthesizes 
$M_{\text{Ni}} < 0.1M_\odot$ \citep{2016ApJ...821...38S}.
The  ${}^{56}$Ni luminosity is 
\begin{equation}
    L(t) = 1.45 \times 10^{43} \bigg( \frac{M_{\text{Ni}}}{M_{\odot}} \bigg) e^{-t/\tau_{0}} \text{erg s}^{-1}
\end{equation}   
where  $\tau_0 = 111.3$ \citep{1994ApJS...92..527N}. The time required for the luminosity to fade below 
a threshold luminosity, L, is
\begin{equation}
    t_{fade} = 3.2 + 0.7\log \Bigg[ \bigg( \frac{10^4 L_\odot}{L} \bigg) \bigg( \frac{M_{\text{Ni}}}{0.1M_\odot} \bigg ) \Bigg] \text{ yr}.
\end{equation}
This neglects $\gamma$-ray escape, which would lead to a more rapid decline in luminosity \citep{2014ApJ...792...10S}. For the typical Ni masses produced in a Type II-P SN of $M_{\text{Ni}} < 0.1M_\odot$ \citep{2016ApJ...821...38S}, the time
scale for the radioactive decay luminosity to be less than the progenitor luminosity is roughly 3-4 years. For example,
SN~1987A had a luminosity of $\sim 30000L_\odot$ ($\sim 3000L_\odot$) after 1000 (2000)~days and $M_{\text{Ni}} = 0.071\pm0.003M_\odot$ \citep{2014ApJ...792...10S}.  We will estimate the required $M_{\text{Ni}}$ masses empirically by scaling the SN~1987A V-band light curve from \cite{2014ApJ...792...10S} to the LBT data.  While we phrase the result in terms of $M_{\text{Ni}}$, the emission at these late times should be from $^{44}$Ti.

\subsection{Circumstellar Medium (CSM) Shock Interactions}

Following the SN, the expanding ejecta will shock heat any pre-existing circumstellar medium (CSM), exciting optical emission lines
which contribute to the late time luminosity (e.g., \citealt{1982ApJ...258..790C}). For a spherically symmetric $\rho \propto 1/r^2$ pre-SN wind, the optical shock luminosity is
\begin{equation*}
     L_s \simeq \epsilon \frac{1}{2}\dot{M}\frac{v_s^3}{v_w}
\end{equation*}
\begin{equation}
    \label{eqn:CSM}
    \simeq 53000 \left( { \frac{\epsilon}{0.1}} \right) \Bigg( \frac{\dot{M}}{10^{-6}M_\odot\text{yr}^{-1}} \Bigg) \Bigg( \frac{v_s}{4000 \text {km~s}^{-1}} \Bigg)^3
    \Bigg( \frac{10 \text{km~s}^{-1}}{v_w} \Bigg) L_\odot 
\end{equation}
where $v_s$ is the shock speed, $\dot{M}$ is the wind mass loss, $v_w$ is the wind speed and $\epsilon$ is the fraction of
the luminosity emitted in the optical.  For luminosities comparable to the progenitor or lower, the inertia of the swept up CSM is only modestly slowing the shock, so the ability of the shock to support late time emission is largely controlled by
the efficiency $\epsilon$ of converting the shock energy into optical emission.  The physics of $\epsilon$ is very complex,
since it depends on the balance of emission mechanisms and the radiative efficiency. For our standard results we will simply
match the R-band luminosity ($\nu L_\nu$) assuming $\epsilon = 0.1$, $v_s = 4000$~km/s and $v_w=10$~km/s.  We chose $\epsilon$
and $v_s$ to match \citet{2022MNRAS.517.4151C} but use $v_w=10$~km/s (instead of 45~km/s) 
to match the assumption of the RSG mass loss models we consider later as well as most $\dot{M}$ estimates from X-ray and radio studies.
We focus on the R band light curve since it contains the strong H$\alpha$ emission line. In our sample, CSM interactions have been invoked for
 SN~1993J (e.g., \citealt{1993ApJ...419L..73S}, \citealt{1994ApJ...431L..95L}, \citealt{1995ApJ...455..658S}, \citealt{1996ApJ...461..993F}), SN~2002hh (e.g., \citealt{2006ApJ...641.1029C}), SN~2004dj (e.g., \citealt{2006ApJ...641.1029C}, \citealt{2012ApJ...761..100C}), SN~2011dh (e.g., \citealt{2012ApJ...752...78S}, \citealt{2014ApJ...785...95M}), SN~2013ej (e.g., \citealt{2017ApJ...834..118M}), and SN~2017eaw (e.g. \citealt{2020ApJ...900...11W}). These $\dot{M}$ estimates are given
 in Table~\ref{tab:analysis_results}.

\begin{figure}
    \centering
    \includegraphics[width = \columnwidth]{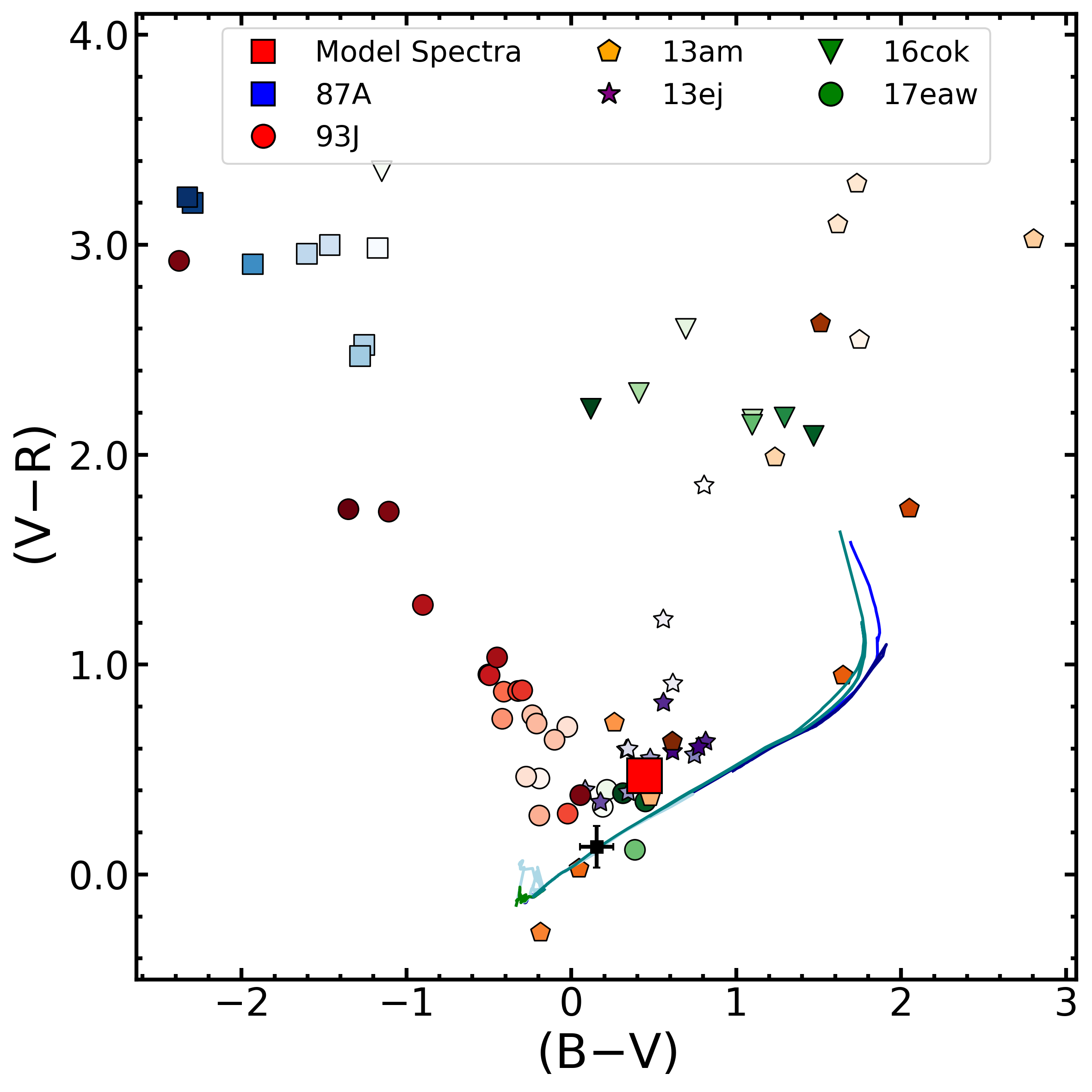}
    \caption{Colors of the late time emission compared to the model CSM interaction spectra (red square) from \citet{2022A&A...660L...9D} and SN~1987A (blue squares) from \citet{2019ApJ...886..147L}. Darker colors represent later time epochs from the peak. The solid lines are Solar metallicity PARSEC isochrones with ages from $\log_{10}$(Age) = 6.6 to 7.2. The black cross is the mean, near-peak, extinction corrected colors for 5 of the SNe. The size of the cross is for visibility - the scatter in the colors is less than the size of the symbol.}
    \label{fig:CSM_color_comp}
\end{figure}

As illustrated by the theoretical models of \citet{2022A&A...660L...9D} or the late time spectra of SN~1993J from
\citet{Matheson2000}, the optical CSM emission is dominated by emission lines,
so the color of the emission should distinguish it from continuum emission like dust echoes and shock heated secondaries.
Fig.~\ref{fig:CSM_color_comp} shows the colors of their model with 1 M${}_{\odot}$ of ejecta travelling at $10^4$km~s${}^{-1}$, which provides
a reasonable match to the \citet{Matheson2000} spectra of SN~1993J $976$~days after peak, as compared to
the Solar metallicity \verb'PARSEC' (\citealt{2012MNRAS.427..127B}, \citealt{2013MNRAS.434..488M}) isochrones with 
ages of $10^{6.6}$ to $10^{7.2}$~years.  We also show the colors of the emission from the shock heated ring in
SN~1987A using the fluxes from \citet{2019ApJ...886..147L} over the time period from 3200 to 11000 days after peak.  In SN~1987A,
the emission peaks nearly 22 years after the explosion, with B, V, R luminosities of roughly $\sim 60L_\odot$,
$\sim 10L_\odot$ and $\sim 160 L_\odot$, respectively, that are orders of magnitude less than the luminosity scales of the
progenitor stars shown in Fig.~\ref{fig:prog}.  This particular color combination, B$-$V and V$-$R, appears to be
a good combination for identifying systems dominated by line emission because the R band contains the strong H$\alpha$
emission, and the B band contains strong O[III] emission while the H$\beta$/O[III] emission lines in the V band
are generally weaker.  This should give CSM emission blue B$-$V colors and red V$-$R colors which are difficult
for less line dominated spectra to mimic, as seen in Fig.~\ref{fig:CSM_color_comp}.

\subsection{Dust Echos}

When light from a SN encounters interstellar dust, it can be scattered to reach us with a light travel time delay. The detailed properties of light echos are dependent on multiple parameters such as dust location, geometry, and scattering optical depth (e.g., \citealt{1986ApJ...308..225C}, \citealt{2003AJ....126.1939S}, \citealt{2003ApJ...582..919L}, \citealt{2005Natur.438.1132R}, \citealt{2011ApJ...732....2R}). 
We can roughly characterize the luminosity as
\begin{equation}
    \label{eqn:echo}
    L = \frac{E_{rad}\tau_{dust}}{t_{now}}
\end{equation}
where $E_{rad}$ is the energy radiated in the photometric band, $t_{now}$ is the time elapsed since peak, and $\tau_{dust}$ is the fraction of $E_{rad}$ that is scattered to the observer. This model essentially assumes
that the dust is in a shell a distance $c t_{now}$ from the SN that is absorbing fraction $\tau$ of the radiated
energy with light travel times then spreading the observed emission over time $t_{now}$.  Dust echoes have (roughly)
the emission weighted mean spectrum of the SN, weighted by the smoothly varying, direction-dependent scattering opacity of the dust. For our sample, dust echoes are reported for SN 1980K (\citealt{2012ApJ...749..170S}, \citealt{2017MNRAS.465.4044B}), SN 1993J (\citealt{2002ApJ...581L..97S}, \citealt{2003ApJ...582..919L}), SN 2002hh (\citealt{2007ApJ...669..525W}, \citealt{2015MNRAS.451.1413A}), SN 2003gd \citep{2005ApJ...632L..17S}.  For these SNe, the observed light echos are generally bluer than the color of the SN near peak.

We again make the estimates using the R band data, in this case because it generally has smaller uncertainties.
The data are collected using the LBC/Red camera for the R-band while cycling through UBV on the LBC/Blue camera, leading to
shorter exposure times for the other filters.  We estimate the  the total radiated energy $E_{rad}$ using the \citet{2015ApJ...814...63M} light curve models normalized to the peak R-band peak luminosity given in Table~\ref{tab:SN_INFO}.
We then simply estimate $\tau_{dust}$ using Eqn.~\ref{eqn:echo}.  The basic physics of echoes requires
the $1/t_{now}$ decay unless the effective optical depth is increasing with distance from the SN.  
Fig.~\ref{fig:CSM_color_comp} shows the mean, near-peak colors of SN~1993J, 2013am, 2013ej, 2016cok, and 2017eaw
after correcting for extinction.  The scatter in the colors is smaller than the symbol.  We would expect dust
echoes to be modestly bluer than these colors.

\subsection{Engine Driven Emissions}

Another possibility for driving late time emissions is energy injection from a pulsar.  If we assume dipole spin down,
the  luminosity is
\begin{equation}
     L =  \frac{4 \pi^2 I}{P^2 t_s} = 330000 I_{45} P_{10}^{-2} t_4^{-1} L{}_{\odot}
\end{equation}
for moment of inertia $I = 10^{45}$~g~cm$^2$, spin period $P=10 P_{10}$~msec, and spin down time
$t_s = 10^4 t_4$~years.  The spin down time is related to the magnetic field $B=10^{12} B_{12}$~gauss by
\begin{equation}
    t_s = P \dot{P}^{-1} = 13000 B_{12}^{-2} P_{10}^2~\hbox{year}.
\end{equation}
For the Crab pulsar ($P=0.033$~s, $t_s=2600$~years), $L=1.1 \times 10^5L_\odot$.  However, essentially
none of this energy is converted into visible radiation at late times. For the optical magnitudes of the Crab from \citet{2009A&A...504..525S}, a distance of $2$~kpc \citep{2008ApJ...677.1201K} and an extinction of $E(B-V) \simeq 0.4$~mag \citep{2019ApJ...887...93G},
the UBVR luminosities of the Crab are $<10L_\odot$.  Magnetar models (e.g., \citealt{2010ApJ...717..245K}), assume that
the spin down energy is fully thermalized and that the amount radiated is regulated by the balance 
between the expansion and diffusion times. We are considering the emission at late times in the nebular
phase, where the prediction of these models is that the luminous emissions are strongly suppressed and
we would expect properties more like the Crab.  For this reason, we do not consider engine driven
emissions further.

\subsection{Binary Companion Ejecta Interaction}

When the shock from a SN hits a binary companion, it heats and inflates its envelope, which leads to a period of enhanced luminosity (\citealt{1975ApJ...200..145W}, \citealt{1981ApJ...243..994F}, \citealt{2000ApJS..128..615M}, \citealt{2003astro.ph..3660P}, \citealt{2007PASJ...59..835M}, \citealt{2014ApJ...792...66H}, \citealt{2018ApJ...864..119H}, \citealt{2021MNRAS.505.2485O}). 
We use the numerically calibrated analytic model of \citet{2021MNRAS.505.2485O} to constrain the companion mass ($M_c$), and the ratio of its radius to the orbital separation ($R_c / a$). The total energy injected into the companion from the outburst is
\begin{equation}
E_{heat} = p  E_{expl}   \tilde{\Omega}
\end{equation}
where $p \simeq8-10\%$ is the energy injection efficiency, $E_{expl}=10^{51}$erg is the explosion energy, and 
\begin{equation}
\tilde{\Omega} = \frac{1}{2} \Bigg[
1 - \sqrt{1 - \Bigg( \frac{R_{c}}{a} \Bigg)^{2}} \Bigg]
\end{equation}
is the fractional solid angle subtended by the companion. The maximum luminosity of the companion is 
\begin{equation}
    L_{max} = \frac{4 \pi G M_c c}{\kappa_{fit}}
    \label{eqn:L_max}
\end{equation}
where $\kappa_{fit}$ is a fitting function from Ogata et al. (2021) for the average opacity at the bottom of the companion's convective layer, $G$ is the gravitational constant, $c$ is the speed of light, and $M_c$ is the companion mass. Thus, the maximum luminosity is determined by the companion mass. The timescale over which the companion maintains this luminosity is,
\begin{equation}
    \tau_{infl} = \alpha \frac{E_{heat}}{L_{max}}
    \label{eqn:Infl_T}
\end{equation}
where $\alpha = 0.18$. Since significant heating of a main sequence (MS) binary companion requires an orbit smaller than the RSG progenitors of Type II SNe, we consider this case only for the Type Ibc SN~2012fh.

\begin{figure*}
    \centering
    \begin{subfigure}[b]{\linewidth}
        \centering
        \includegraphics[width=0.75\textwidth]{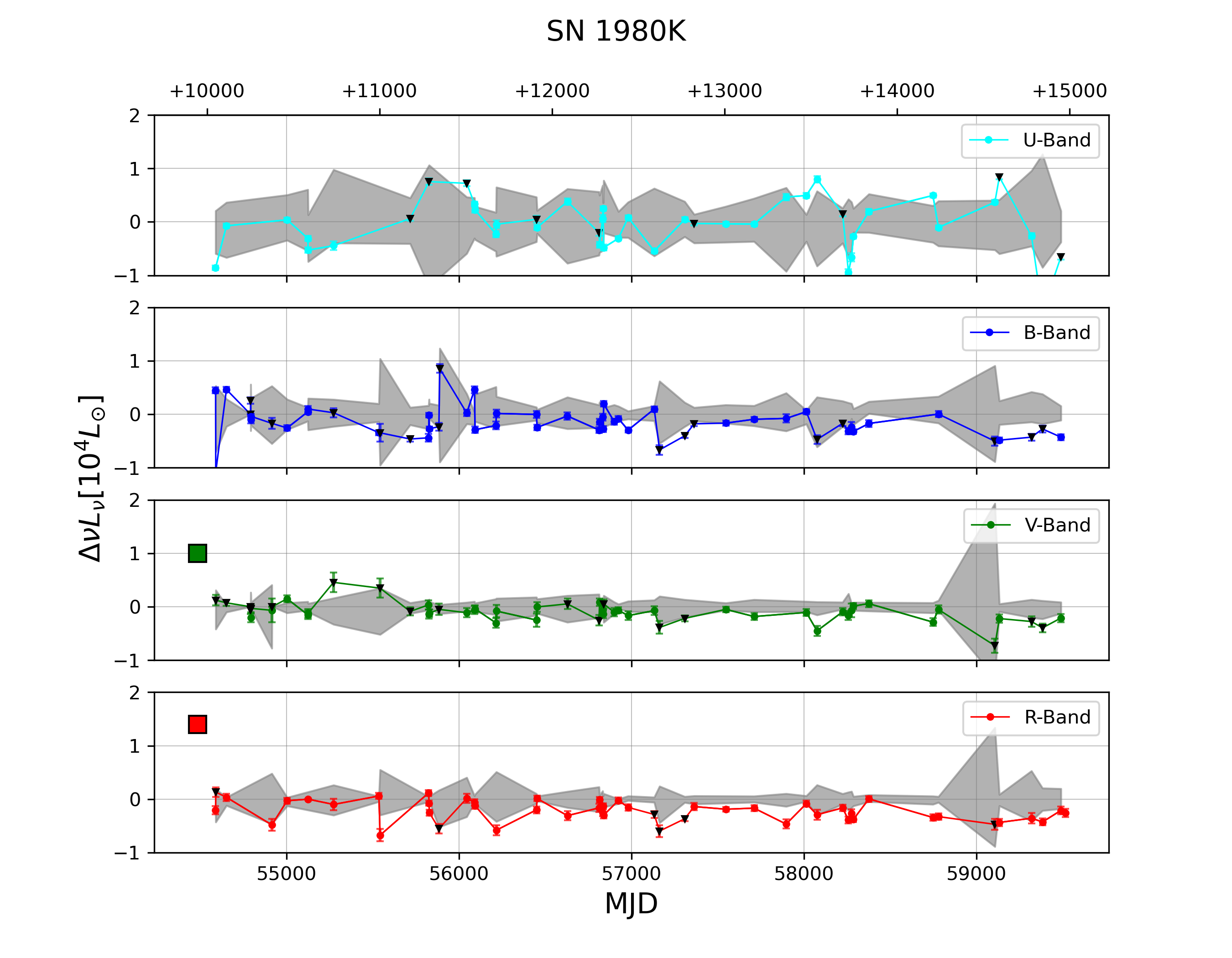}
    \end{subfigure}
    \begin{subfigure}[b]{\linewidth}
        \centering
        \includegraphics[width=0.75\textwidth]{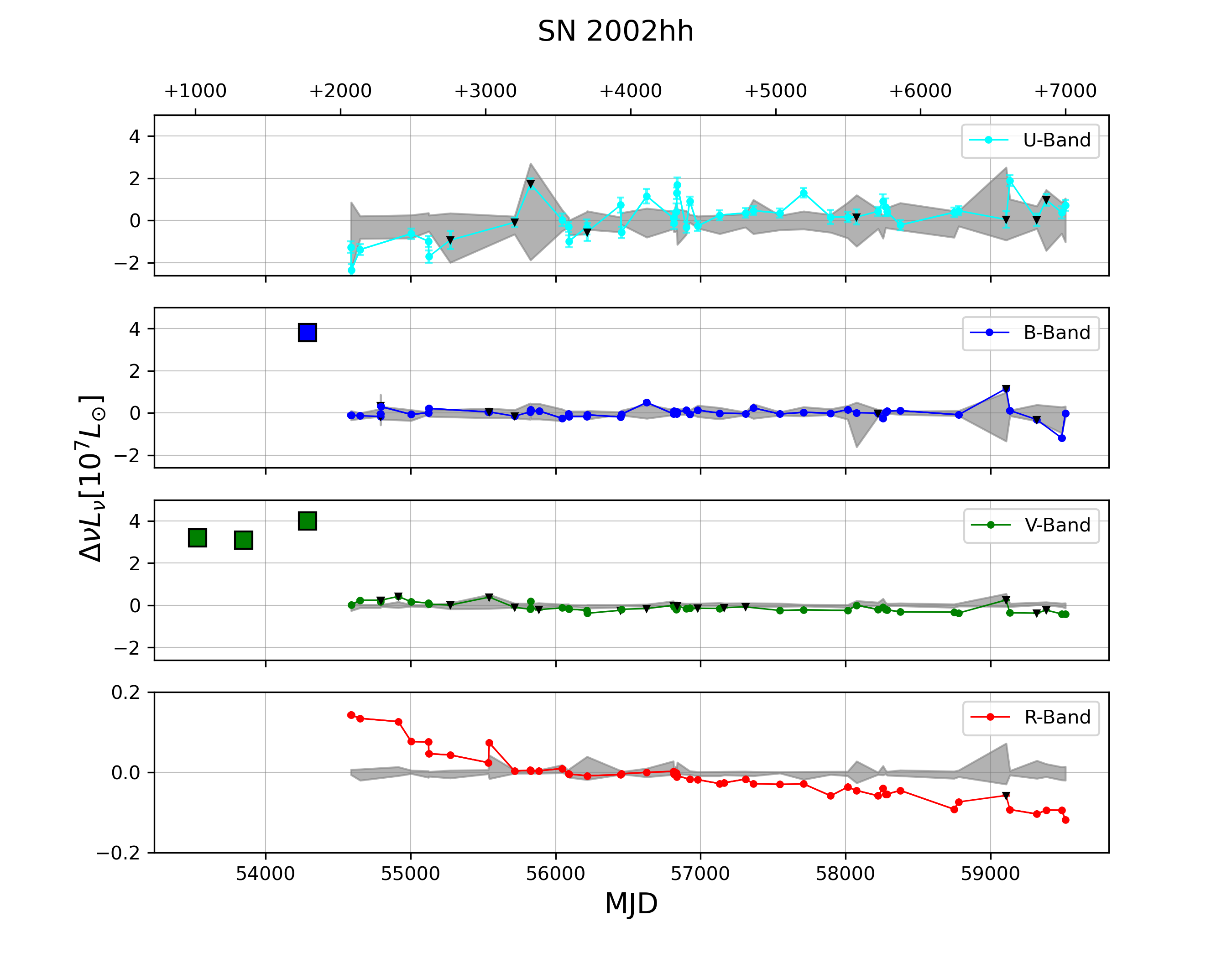}
    \end{subfigure}
    \caption{ LBT UBVR difference imaging light curves for the Type IIP/L SNe with days after peak shown at the top and black triangles for poorer quality epochs as defined in \S~\ref{sec:observation}. The gray region is the 1$\sigma$ scatter about the mean of the 11 comparison light curves. For SNe with pre-SN images, the vertical dashed line is the date of peak. The large squares are the HST photometry from Table~\ref{tab:HST_DATA} where available. The star symbol denotes the epoch at which the linear fits provided in Tables~\ref{tab:SN Decay}~and~\ref{tab:SN Decay Cont.} begin. For SNe which exploded prior to the LBT survey, the entire light curve is fit. Except for 2011dh, 2013am, 2016cok, and 2017eaw where the luminosity is in units of $10^5L_\odot$, 2009hd in units $10^6L_\odot$, and 2002hh in units $10^7L_\odot$, the luminosity is in units of $10^4L_\odot$. The range shown varies between the SNe for clarity. The luminosity is relative to the luminosity in the reference image, which for SNe with pre-explosion imaging is the luminosity of the progenitor.}
    \label{fig:MultiStackIIP/L}
\end{figure*}

\begin{figure*}
    \centering
    \begin{subfigure}[b]{\linewidth}
        \centering
        \includegraphics[width=0.8\textwidth]{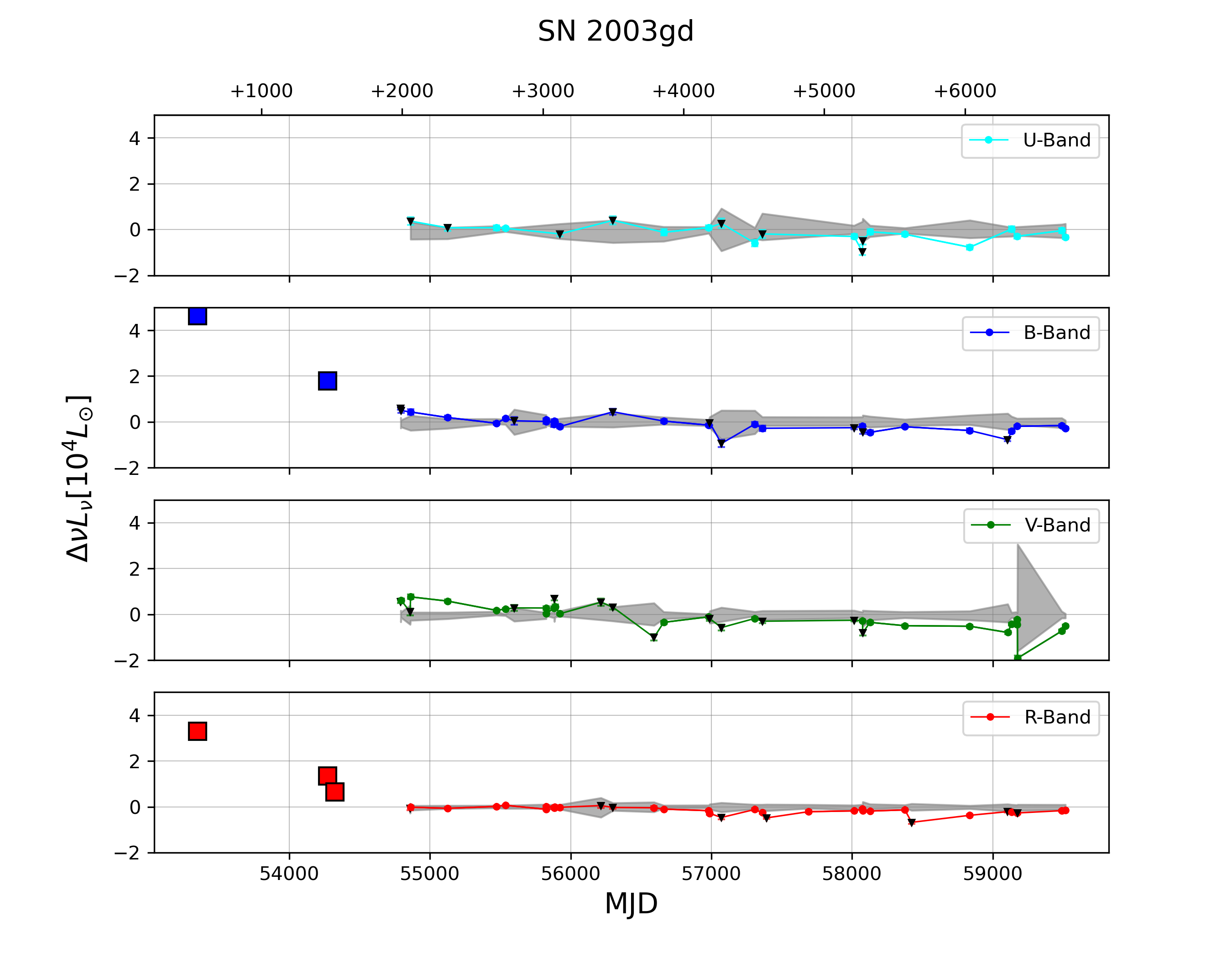}
    \end{subfigure}
    \begin{subfigure}[b]{\linewidth}
        \centering
        \includegraphics[width=0.8\textwidth]{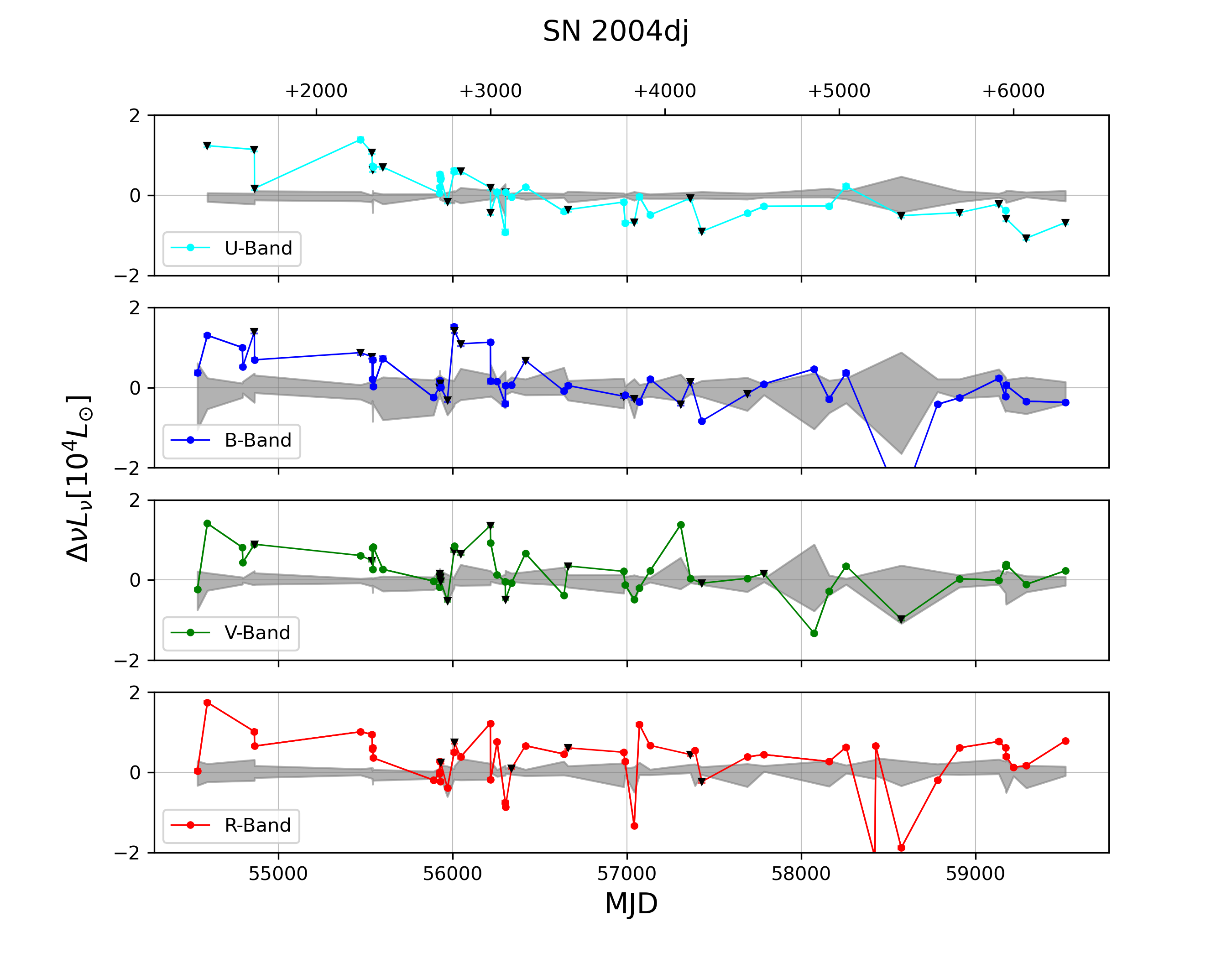}
    \end{subfigure}
    \caption{Type IIP/L (cont.)}
    \label{fig:IIP/Lcont}
\end{figure*}

\begin{figure*}
    \centering
    \begin{subfigure}[b]{\linewidth}
        \centering
        \includegraphics[width=0.8\textwidth]{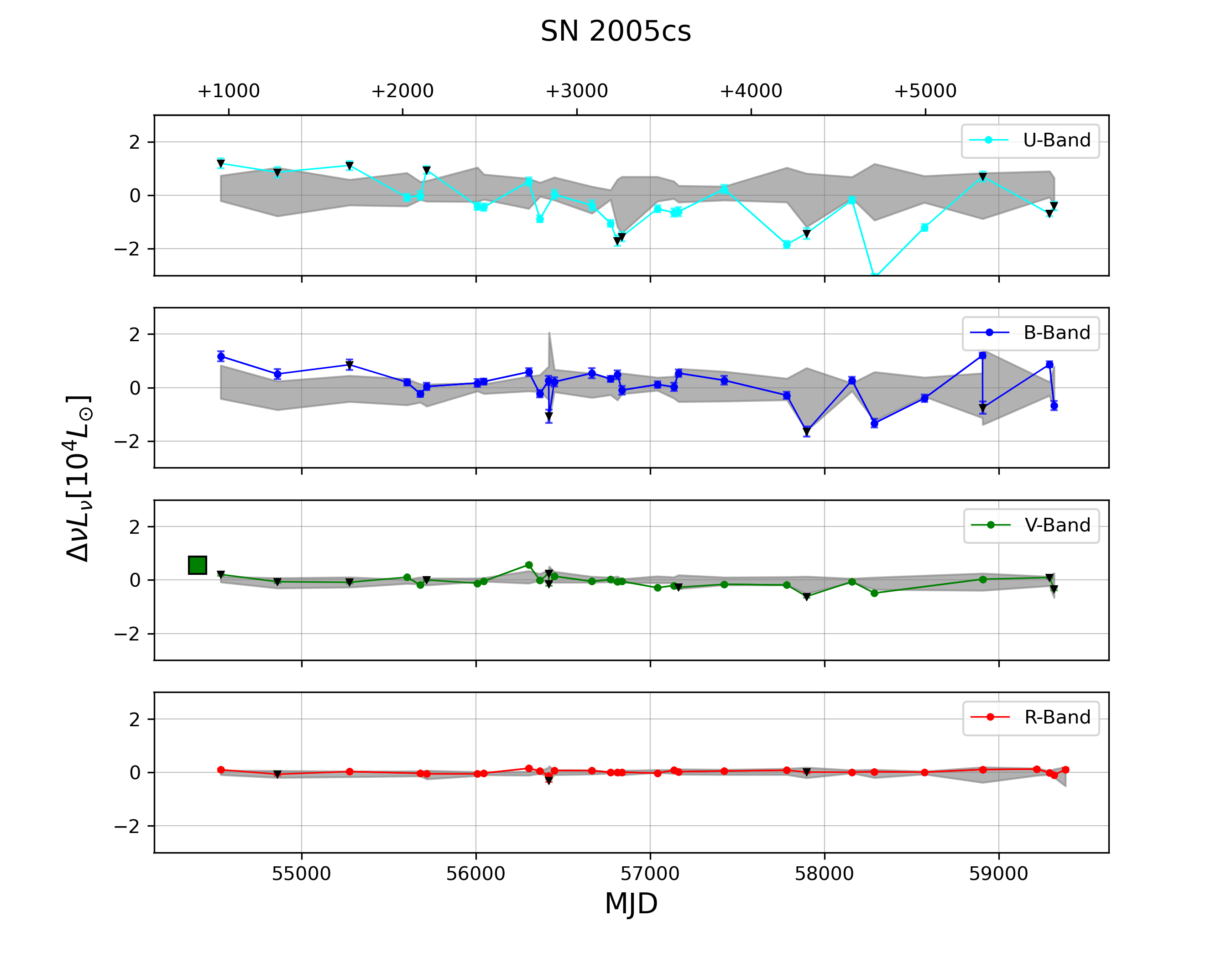}
    \end{subfigure}
    \begin{subfigure}[b]{\linewidth}
        \centering
        \includegraphics[width=0.8\textwidth]{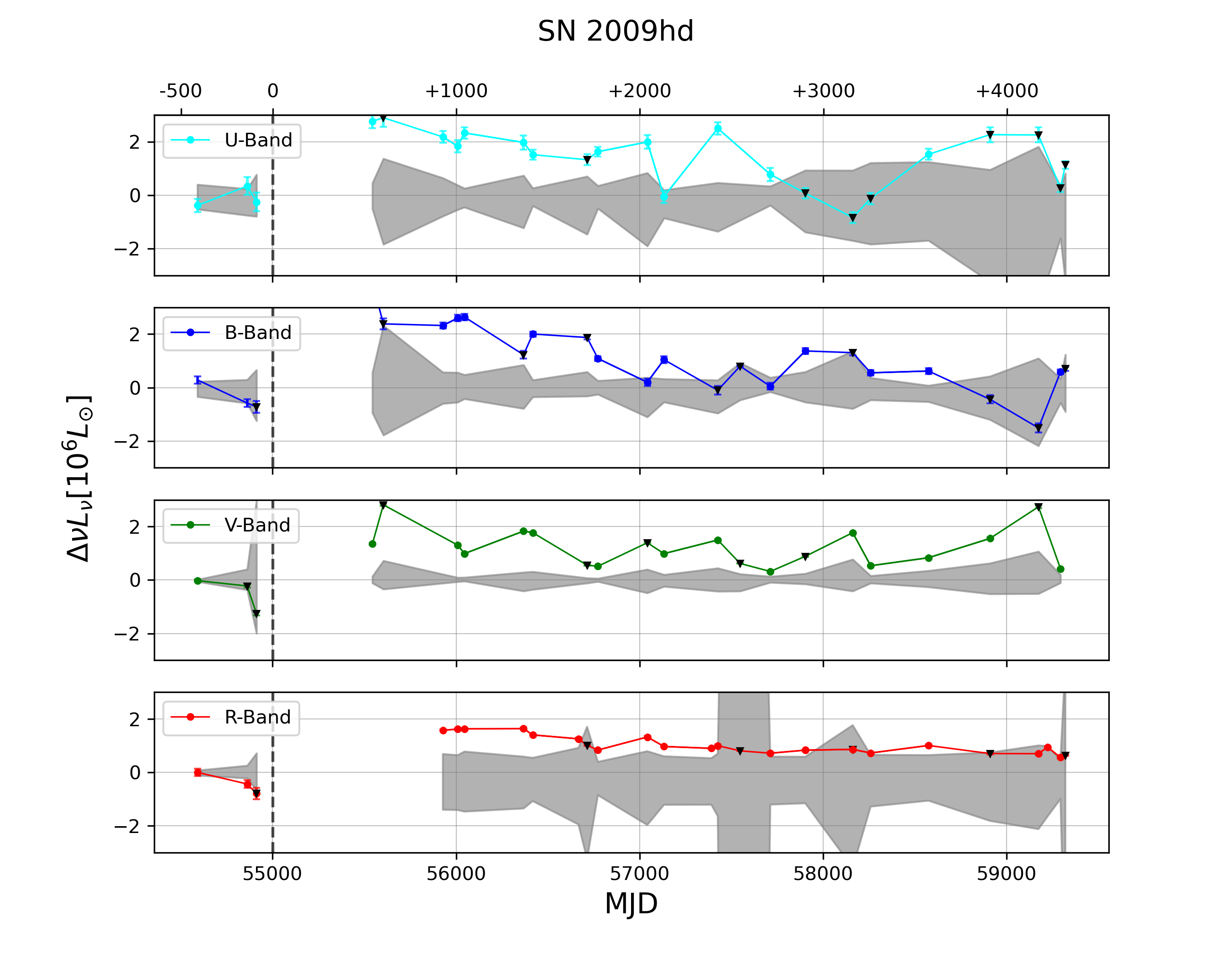}
    \end{subfigure}
    \caption{Type IIP/L (cont.)}
    \label{fig:IIP/Lcont2}
\end{figure*}
\begin{figure*}
    \centering
    \begin{subfigure}[b]{\linewidth}
        \centering
        \includegraphics[width=0.8\textwidth]{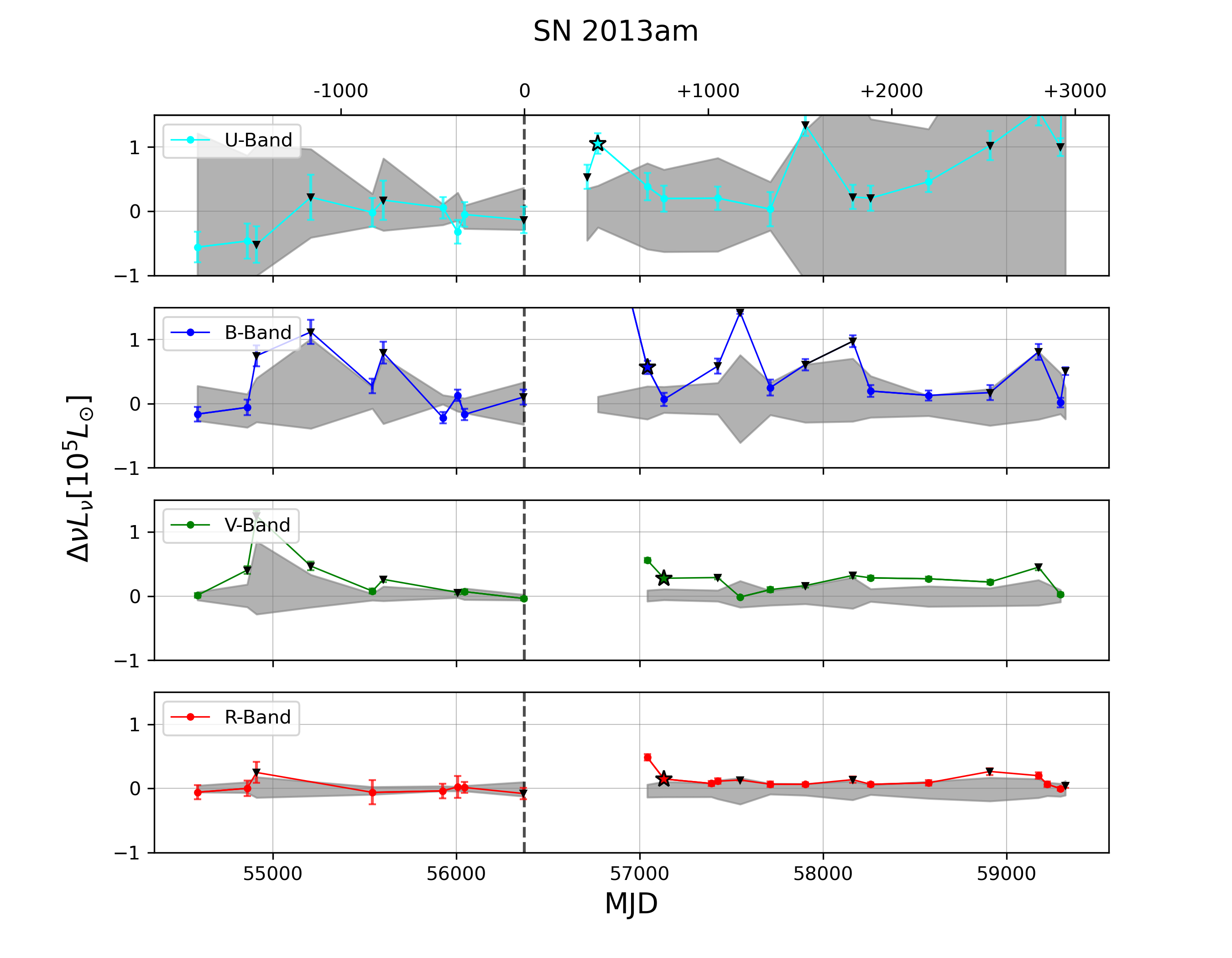}
    \end{subfigure}
    \begin{subfigure}[b]{\linewidth}
        \centering
        \includegraphics[width=0.8\textwidth]{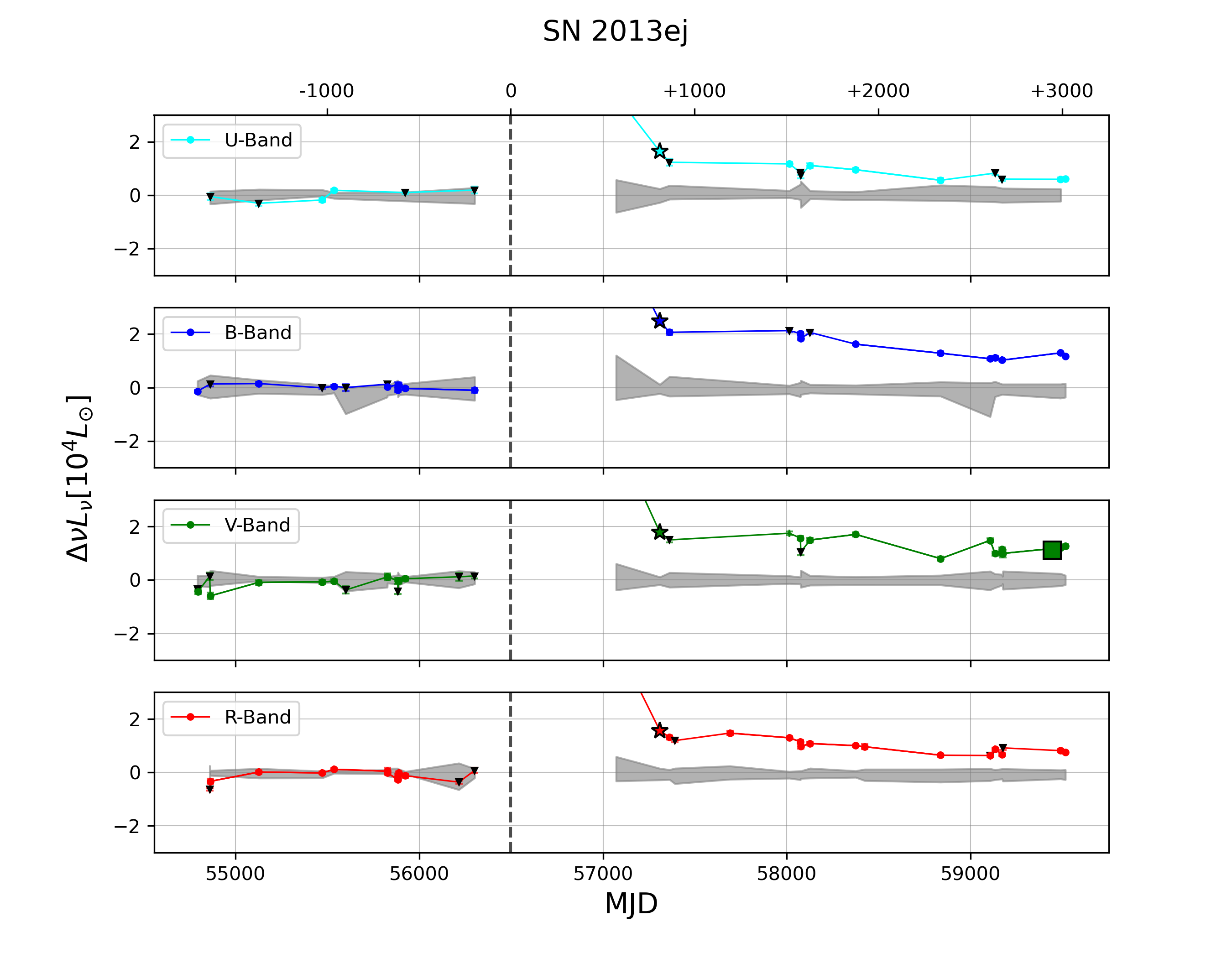}
    \end{subfigure}
    \caption{Type IIP/L (cont.)}
    \label{fig:IIP/Lcont3}
\end{figure*}

\begin{figure*}
    \centering
    \begin{subfigure}[b]{\linewidth}
        \centering
        \includegraphics[width=0.8\textwidth]{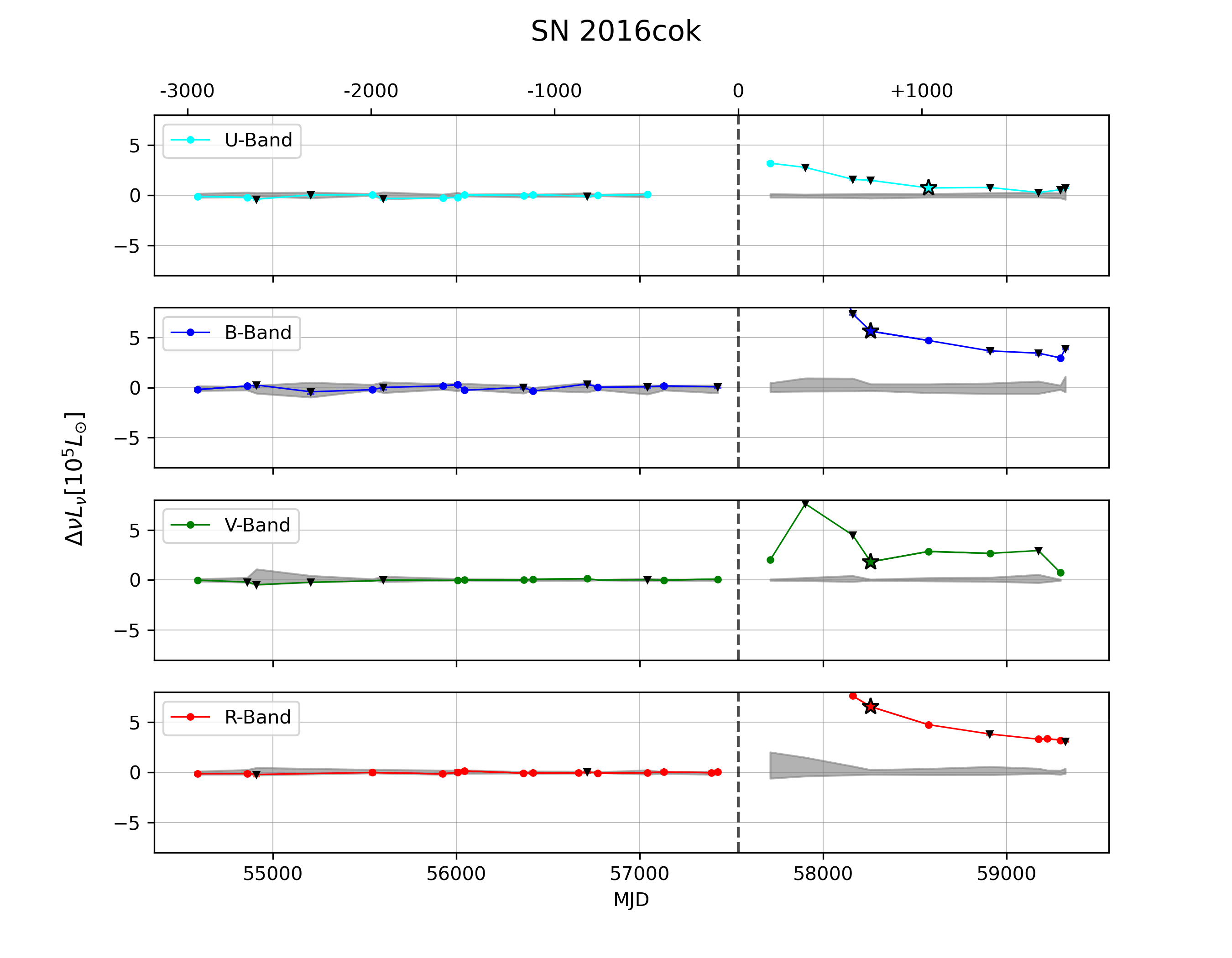}
    \end{subfigure}
    \begin{subfigure}[b]{\linewidth}
        \centering
        \includegraphics[width=0.8\textwidth]{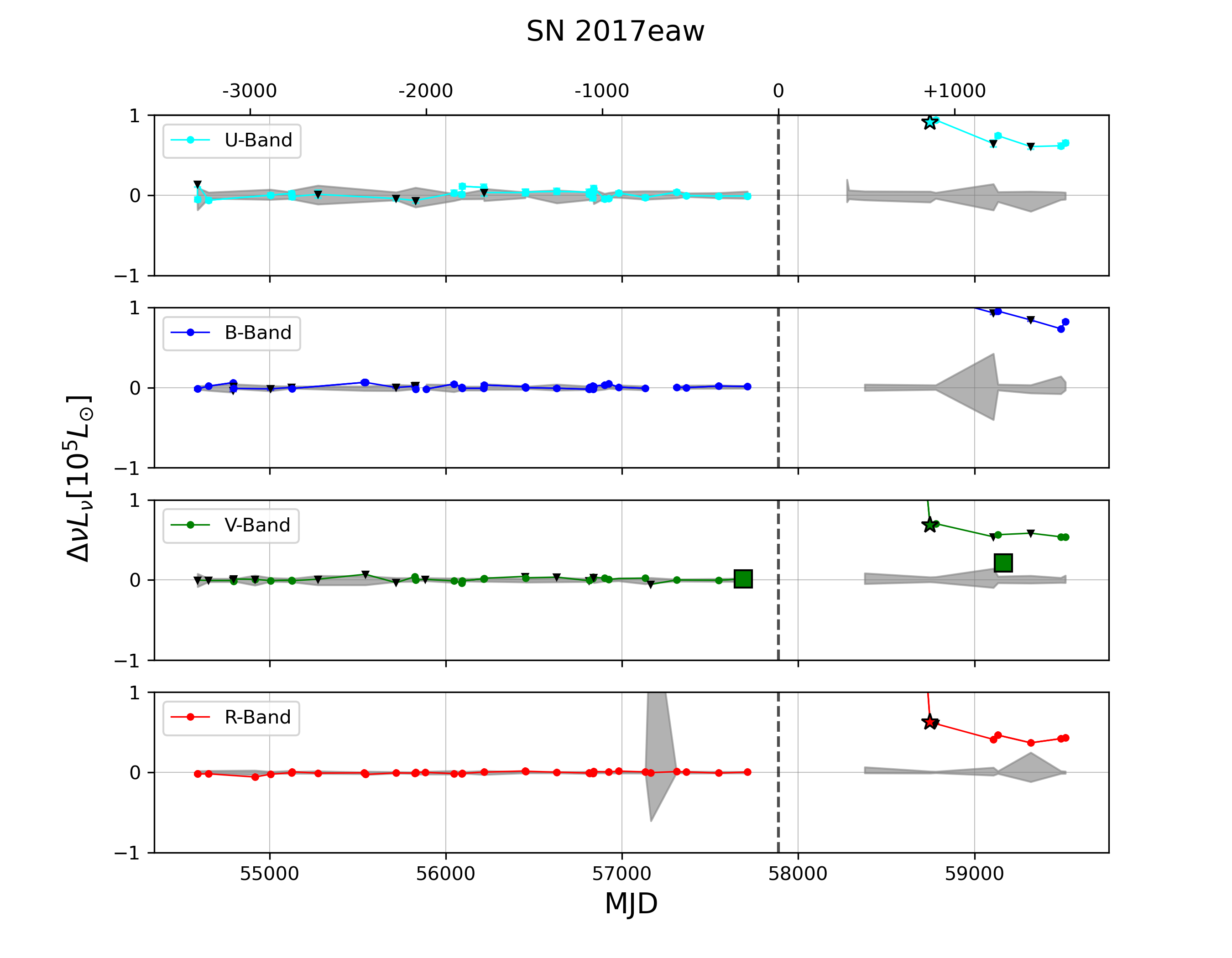}
    \end{subfigure}
    \caption{Type IIP/L (cont.)}
    \label{fig:IIP/Lcont4}
\end{figure*}

\begin{figure*}
    \centering
    \begin{subfigure}[b]{\linewidth}
        \centering
        \includegraphics[width=0.8\textwidth]{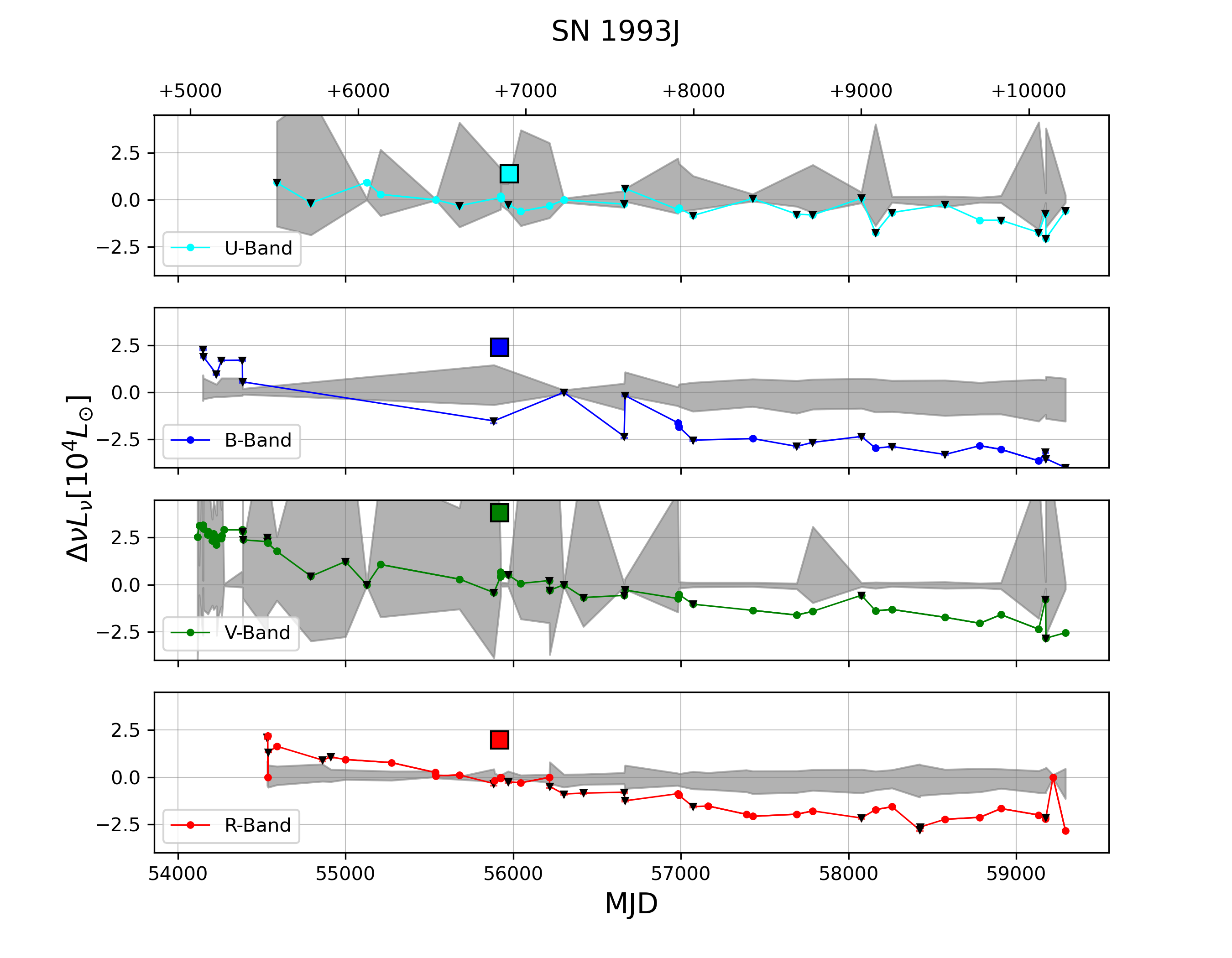}
    \end{subfigure}
    \begin{subfigure}[b]{\linewidth}
        \centering
        \includegraphics[width=0.8\textwidth]{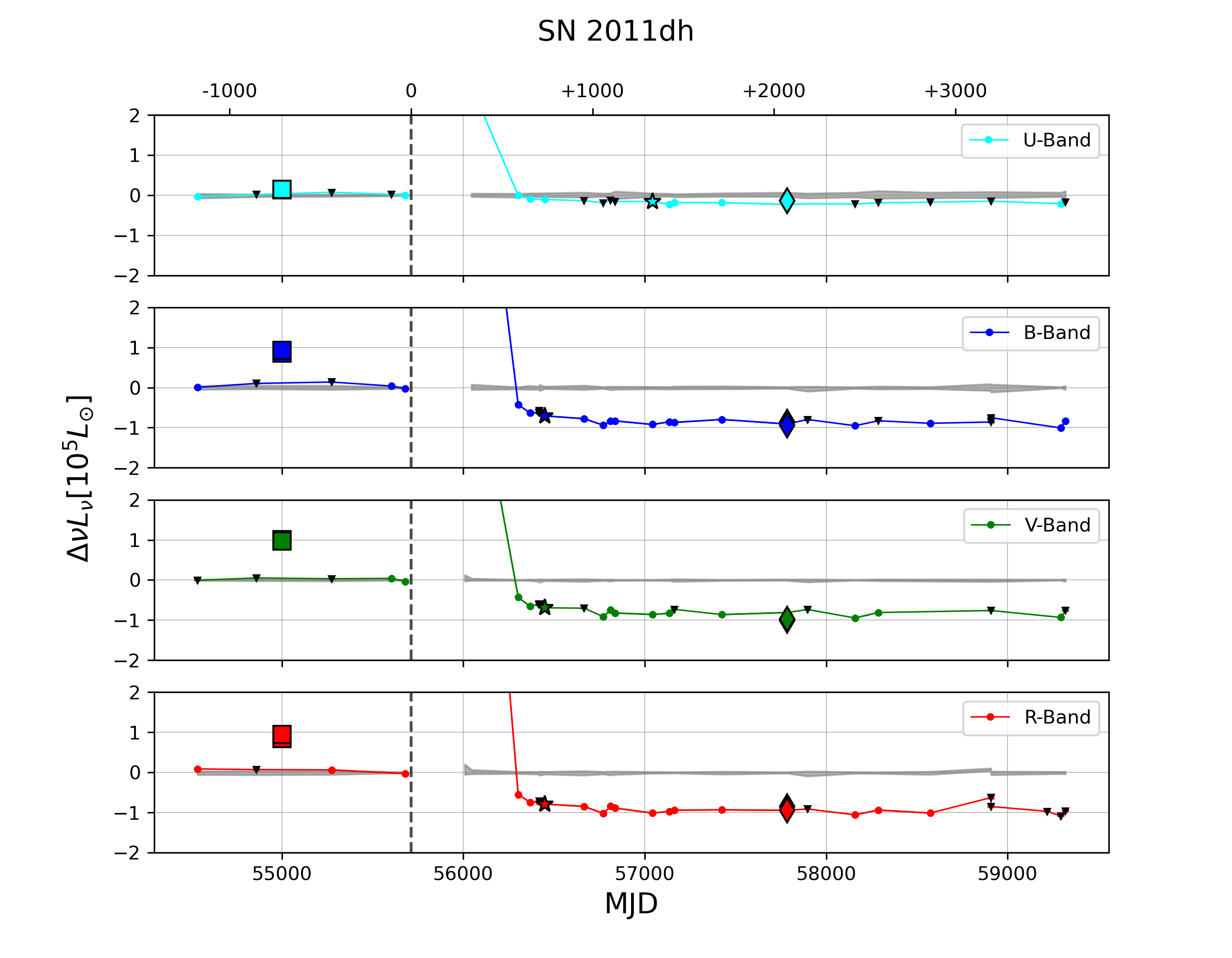}
    \end{subfigure}
    \caption{LBT UBVR difference imaging light curves for the Type IIb SNe. The large squares are the HST photometry from Table~\ref{tab:HST_DATA} where available. For SN~2011dh the pre-SN HST observations occurred at an earlier epoch and are shifted to the date shown. For SN~2011dh, the diamonds are the inverse of HST photometry at the epoch we perform the SED fit as described in \S~\ref{sec:proc_con}.}
    \label{fig:MultiStackIIb}
\end{figure*}

\begin{figure*}
    \centering
    \begin{subfigure}[b]{\linewidth}
        \centering
        \includegraphics[width=0.8\textwidth]{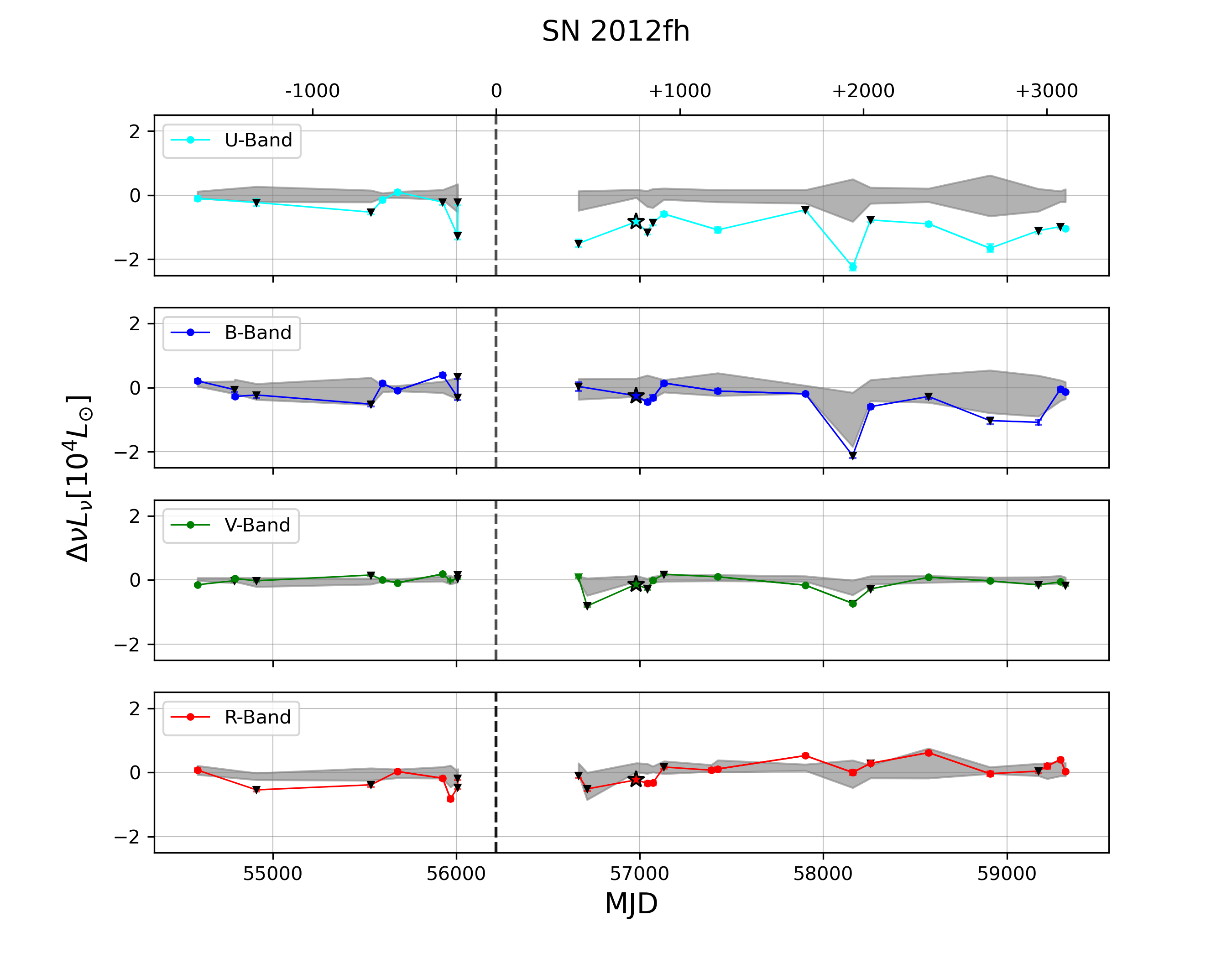}
    \end{subfigure}
    \caption{LBT UBVR difference imaging light curves for the Type Ib/c SN}
    \label{fig:MultiStackIb/c}
\end{figure*}

\section{Discussion}\label{sec:Discussion}

Table~\ref{tab:SN_INFO} summarizes the SN we consider, and the Appendix~\ref{appendix} has notes on the individual sources. After dropping SN~2009hd, there are 12 sources in total; 9 IIL/P, 2 IIb, and 1 Ib/c, of which 4, 1, and 1 have pre-SN LBT observations. Figures~\ref{fig:MultiStackIIP/L}--\ref{fig:MultiStackIb/c} show the LBT light curves of the SN grouped by type and with the epoch of explosion marked if it occurred after the start of the LBT project.  
Tables~\ref{tab:SN Decay}~and~\ref{tab:SN Decay Cont.} give the results of the linear fits to the late time light curves of each SNe and its grid of comparison points where the fits extend from the epoch marked in the figure through the end of the light curve. We include fits for SN~2009hd even though we will not discuss it further.

The luminosities of the SNe in these light curves are all relative to the luminosity of the source in the reference image. For the SNe with pre-SN LBT data, the observed light curve is thus any present day emission minus the luminosity of the progenitor. For the SN with only late time LBT data, it is the present day emission minus the mean luminosity of the SN is the reference image flux. The aperture photometry luminosity of the SNe on the reference image is reported in Tables~\ref{tab:SN Decay} and~\ref{tab:SN Decay Cont.}, but they should be generally regarded as limits due to the effects of crowding. Figures~\ref{fig:MultiStackIIP/L}--\ref{fig:MultiStackIb/c} include the HST luminosities from Table~\ref{tab:HST_DATA}.  The offset between the differential luminosity light curve and the HST luminosity is
the luminosity in the reference image.   Where the HST data is available, the actual light curve is simply the
differential light curve shifted upwards in luminosity to pass through the HST data.  What we see in general is
that the necessary shifts are on the same scale as the changes in luminosity.  The exception is SN~1980K, where
the HST luminosity is roughly an order of magnitude larger.  SN~1980K is also the only case where we can clearly
see the SN in the LBT data at these late times.  

For our standard analyses we use the image subtraction light curve if the linear fit estimate of the luminosity of the last epoch is positive.  If the linear fit estimate of the luminosity of the last epoch is negative, we rescale the final luminosity to zero and
use $L_{SN}' = L_{SN} - \beta_{SN}(t_{last}-t_0)$ where $t_{last}$ the time of the last epoch.  All of these latter
systems are SNe without pre-explosion LBT images with the exception of SN~2011dh.  This still means that our estimates of the
$^{56}$Ni mass, wind $\dot{M}$, and dust optical depth $\tau$ will be underestimates, so we discuss the consequences of further increases in the luminosities due to the (remaining) uncorrected flux in the reference image.   We will also provide estimates
based on the slope $\beta_{SN}$.  For radioactive decay we can make a direct estimate of the $^{56}$Ni mass by matching the
slope of the LBT light curve to that of SN~1987A.  For the other two emission mechanisms we make a ``standardized'' estimate using the luminosity $\beta_{SN} \Delta t$, the drop in luminosity over $\Delta t =10$~years.  It is standardized in the sense that the time baselines for the most recent SN are much shorter than the full time baseline of the LBT data.

Of the 9 Type~IIP/L systems, 8 show continued late time emission, particularly in the V and R bands. One of the Type~IIb SN, SN~1993J, shows continued fading.  Only the Type~IIP SN~2005cs, the Type~IIb SN~2011dh and the Type~Ibc SN~2012fh do not show significant evidence for continued emission. For the systems with pre-LBT images, all but SN~2011dh and possibly SN~2012fh have a present day luminosity in excess of the progenitor.  In the next sections we first discuss the progenitor of SN~2011dh, and then the potential
sources of the continued emission for the other systems.

\begin{figure}
     \centering
   \includegraphics[width=0.45\textwidth]{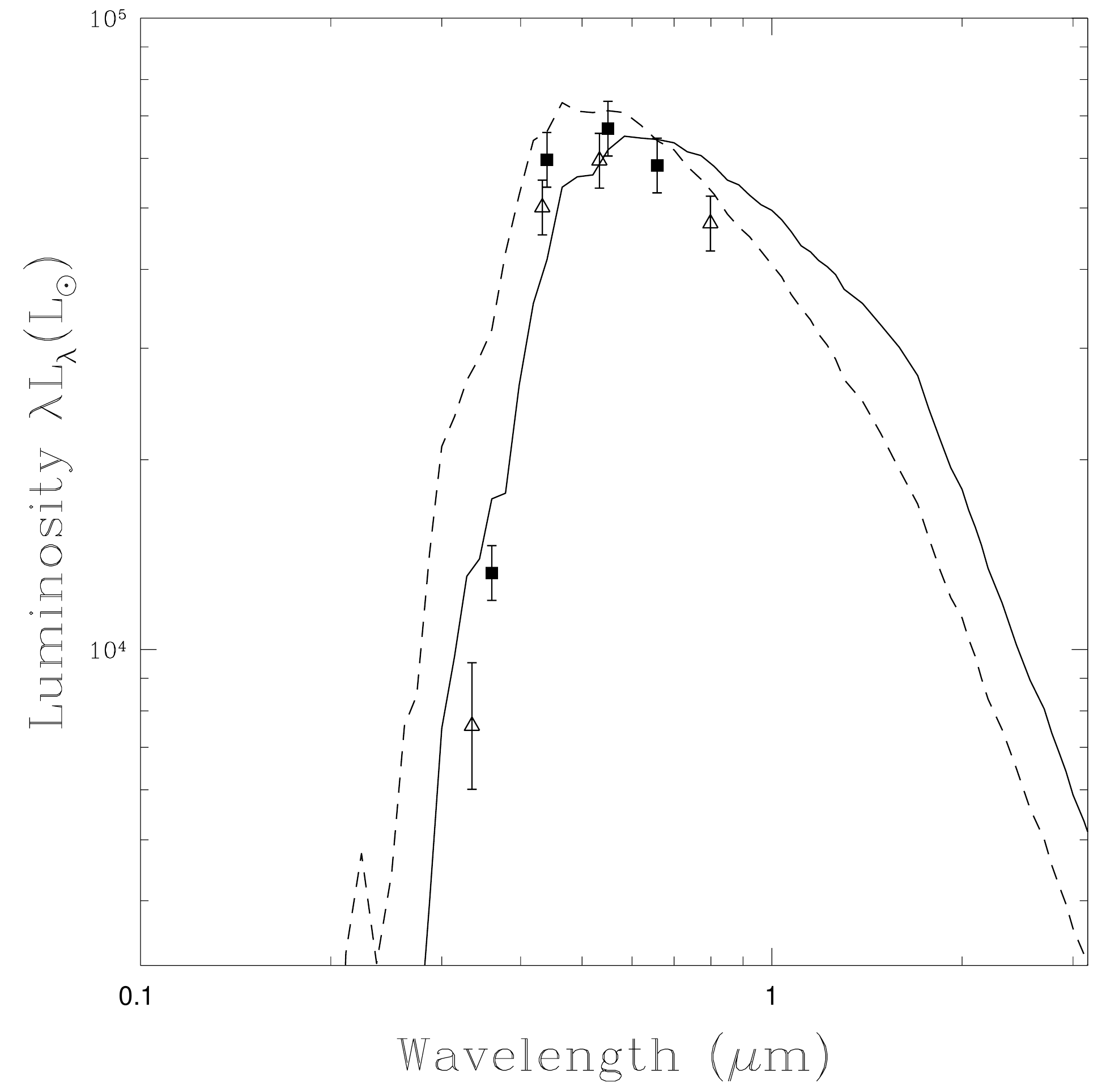}
   \caption{Spectral energy distribution models for the LBT (solid squares) and HST (open triangles) photometry of the
      progenitor of SN~2011dh fitting the temperature, luminosity, and extinction.  The solid curve fits all of the points, while the dashed  curve
      does not include the U/F336W data in the fits.
   }     \label{fig:11dhsed}
\end{figure}

\begin{table}
    \centering
    \begin{tabular}{lccrc}
        \hline
       SN & Date & Band & \multicolumn{1}{c}{Luminosity} & Ref. \\
        & (MJD) &  &\multicolumn{1}{c}{($10^4L_\odot$)} & \\
        \hline
        \hline
        SN~2003gd & 52511 & V & $0.26$ & 1 \\
        SN~2005cs & 53380 & V & $<0.32$ & 2 \\
        SN~2017eaw & 57687 & V & 0.16 & 3 \\
        \hline
       SN~1980K & 54484 & V & $1.0$ & 4\\
         & 54484 &R & $1.4$  &  4\\
        SN~1993J & 	55976 & U & $1.39$ & 5   \\
        & 55919 & B &  $2.39$ & 5\\
        & 55919 & V & $3.81$ & 5 \\
        & 55919 &R & $1.97$ & 5\\ 
        SN~2002hh & 53630 &V & $3200$ & 6 \\
        & 53848 & V & 3100 & 6  \\
        & 54290 & B & 3800 & 6  \\
        & 54290 & V & 4000 & 6 \\
        SN~2003gd & 53347 & B & 4.63 & 7 \\
        & 53347 & R & 3.30 & 7 \\
        & 54272 & B & 1.79 & 8  \\
        & 54272 & R & 1.35 & 8 \\
        & 54323 & R & 0.64 & 6 \\
        SN~2005cs & 54079 & V & 9.66 & 6  \\
        & 54401 & V & 0.56 & 6    \\
       SN~2013ej & 59445 & V & $1.12$ & 9 \\
       SN~2017eaw & 59164 & V & $2.13$ & 9 \\
       
        \hline
    \end{tabular}
    \caption{Existing HST photometry of six SNe in this sample, the horizontal line separates the pre-explosion observations from the post-explosion observations. References: (1) \citet{2004Sci...303..499S}, (2) \citet{2006ApJ...641.1060L}, (3) \citet{2019ApJ...875..136V}, (4) \citet{2012ApJ...749..170S}, (5) \citet{2014ApJ...790...17F}, (6) \citet{2012ApJ...744...26O}, (7) \citet{2005ApJ...632L..17S}, (8) \citet{2009Sci...324..486M}, (9) \citet{2022MNRAS.tmp.3313V} }
    \label{tab:HST_DATA}
\end{table}

\subsection{Progenitor Constraints}\label{sec:proc_con}

As noted earlier, only SN~2011dh, a Type~IIb SNe in NGC~5194, shows the originally expected fading.  The SN was fainter than the
progenitor within $\sim 600$~days, as expected from a light curve dominated by radioactive decay. In a reverse time average of the subtracted light curves, we begin to see an upward trend at times earlier than +1400 days post explosion, so we adopt
the reverse time average of the light curves after day +2000 as the estimate of the progenitor flux.  The principal limitation on our photometry is the calibration because the depths of the LBT data and the SDSS survey are poorly matched (particularly since SDSS only catalogs stars
around the periphery of the galaxy).  A shallower LBT image is needed to improve the calibration.  Nonetheless, the LBT magnitudes
in Table~\ref{tab:11dh_prog} are very similar to those found by \citet{2011ApJ...741L..28V} and \citet{2011ApJ...739L..37M} analyzing the same HST F336W, F435W, F555W, F658N, and F814W images, roughly corresponding to the U, B, V, R and I bands. That they are so similar 
despite the need for a better calibration of the LBT data is a demonstration that it should be straight forward to
obtain very good photometry of SNe progenitors from the LBT data once they have faded. Fig.~\ref{fig:11dhsed} provides an extinction corrected SED.

We fit the data assuming minimum photometric errors of 10\%, a fixed distance of $7.1\pm 1.2$~Mpc to 
match \citet{2011ApJ...741L..28V} and \citet{2011ApJ...739L..37M}
and varying only the luminosity, temperature and foreground extinction, with an extinction
prior of $E(B-V) = 0.05 \pm 0.05$~mag.  This includes both the Galactic extinction and any
contribution from the host. For the LBT data the best fit has $\chi^2=21.0$ for
two degrees of freedom, with 
$L_* = 10^{5.00 \pm 0.04} L_\odot$, $T_* = 5294 \pm 154$~K and
$E(B-V)=0.02\pm 0.02$~mag. The poor
fit is driven by an inability to find a model which is roughly flat in $\nu L_\nu$ for the B, V and
R bands  and then drops rapidly enough to fit the U band.  

We did not fit the HST narrow band H$\alpha$ data (F658N), and for
the \citet{2011ApJ...741L..28V} photometry the best fit has $\chi^2=8.2$ again for two degrees of freedom
with $L_* = 10^{4.98\pm0.04}L_\odot$, $T_*=6312 \pm 357$~K and $E(B-V)=0.10 \pm 0.04$~mag. 
 In their analysis, \citet{2011ApJ...741L..28V} 
adopted a temperature of $T_*=6000$~K and a luminosity of $L = 10^{4.99}L_\odot$ and
a fixed (Galactic) extinction of $E(B-V)=0.04$~mag.
Finally, for the \citet{2011ApJ...739L..37M} values we find $\chi^2=15.2$, $L_* =10^{5.00\pm 0.03}L_\odot$, 
$T_*=6317 \pm 353$~K and $E(B-V)=0.12\pm 0.02$mag. \citet{2011ApJ...739L..37M} found 
$L_*=10^{4.92 \pm 0.20}L_\odot$ and $T_*=6000\pm 280$~K.  Both of the HST fits also struggle
with the rapid drop down to the F336W point, but the impact on the $\chi^2$ is smaller because
of the larger uncertainty for HST F336W compared to the LBT U band.  Using circumstellar
dust instead of foreground extinction did not solve the problem of the poor fits to these data.

If we combine the LBT and \citet{2011ApJ...739L..37M} data (again excluding the narrow band filter)
we get $\chi^2=39.6$ for 7 degrees of freedom, $L_*=10^{4.97\pm0.04}L_\odot$, $T_*=5601\pm226$~K and
$E(B-V)=0.04 \pm 0.04$. If we drop the the U/F336W points, we find $\chi^2=2.1$ for 5 degrees
of freedom with $L_*=10^{4.99 \pm 0.05}L_\odot$, $T_*=6629\pm 324$~K and $E(B-V)=0.08\pm 0.04$~mag.
Fig.~\ref{fig:11dhsed} shows these combined models.
For the adopted distance estimate,
there is an additional uncertainty in the luminosity of $0.15$~dex from the uncertainties
in the distance.  This is likely already included in the \citet{2011ApJ...739L..37M} luminosity uncertainty. All of these models agree on a progenitor luminosity very close to $10^5 L_\odot$
and Fig.~\ref{fig:prog} shows how these luminosity estimates compare to the
progenitor models. The close match to the HST data emphasizes the point in Fig~\ref{fig:13model} that faded progenitors should generally be trivially visible in the LBT data. 

\citet{2019ApJ...883...86M} argues for a late-time emission plateau at 2000-2500~days of roughly $10^4 L_\odot$ 
fading by approximately $(-330 \pm 110)L_\odot$/year, and argues for a dust echo since there are no indications of
the presence of emission lines in their narrowband images.  Such a low level contribution would have little effect
on this analysis since the estimated progenitor luminosity is an order of magnitude higher. Formally,
 we find somewhat steeper decay rates in Table~\ref{tab:SN Decay Cont.}.

\begin{table}
    \centering
    \begin{tabular}{cccc}
        \hline
       Band & LBT & Van Dyk & Maund\\
            & [mag]  & [mag]     &[mag] \\
        \hline
        \hline
        U & $23.18\pm0.07$ & $23.434\pm0.339$ & $23.39\pm0.25$   \\
        B & $22.30\pm0.02$ & $22.451\pm0.005$ & $22.36\pm0.02$    \\
        V & $21.66\pm0.09$ & $21.864\pm0.006$ & $21.83\pm0.04$    \\
        R & $21.42\pm0.04$ & $21.392\pm0.021$ & $21.28\pm0.04$    \\
        I &                & $21.216\pm0.005$ & $21.20\pm0.03$    \\
        \hline
    \end{tabular}
    \caption{Photometry of the progenitor of SN~2011dh. The \citet{2011ApJ...741L..28V} and
      \citet{2011ApJ...739L..37M} magnitudes are the HST F336W, F435W, F555W, F658N and F814W filters.}
    \label{tab:11dh_prog}
\end{table}

\begin{table*}
    \centering
    \begin{tabular}{lrrrrrrcrr}
        \hline
        & \multicolumn{3}{c}{$^{56}$Ni Mass} &  \multicolumn{2}{c}{Mass Loss Rate} &  \multicolumn{2}{c}{Scattering Optical Depth} & \multicolumn{2}{c}{Other $\dot{M}$ Estimates}\\
         & \multicolumn{3}{c}{[M$_\odot$]} & \multicolumn{2}{c}{[$\log_{10}(M_\odot\text{yr}^{-1})$]} & \multicolumn{2}{c}{[$\log_{10}(\tau)$]} &   \multicolumn{2}{c}{[$\log_{10}(M_\odot\text{yr}^{-1})$]} \\
        \multicolumn{1}{c}{SN} & \multicolumn{1}{c}{$10^3L_\odot$}  & \multicolumn{1}{r}{L${}_{SN}$} & \multicolumn{1}{c}{$\beta_{SN}$} & \multicolumn{1}{c}{L${}_{SN}$}  & \multicolumn{1}{c}{$\beta_{SN}\Delta t$} &
        \multicolumn{1}{c}{L${}_{SN}$}  & \multicolumn{1}{c}{$\beta_{SN}\Delta t$} & $\dot{M}$& Ref. \\
        \hline
        \hline
        SN 1980K    &   5.9 & 7.2 & $0.7\pm1.2$ & $-6.65$ & $-7.22$  & $-4.09$ & $-5.10$  & $-4.58$ & (1) \\
        SN 1993J    &   5.9 & 149 & $30.9\pm2.3$  & $-6.02$ & $-6.24$  & $-2.66$ & $-2.88$ & $-4.40$ & (2) \\
        SN 2002hh   &   5.9 & $\sim10^4$ &  $(5\pm0.5)\times10^3 $ & $-4.83$ & $-4.60$  & $-2.98$ & $-3.09$ & $-5.15$ & (3) \\ 
        SN 2003gd   &   5.9 & 23.4 &  $9.8\pm1.1 $  & $-6.99$ & $-7.43$  & $-3.52$ & $-3.40$ \\
        SN 2004dj   &  4.9 & 1.3 & $6.4\pm0.9$  & $-6.67$ & $-7.27$  & $-2.55$ & $-3.17$ &  $-5.60$, $-6.49$ & (3), (4) \\
        SN 2005cs   &   5.6 & 4.9 & $2.3\pm0.3$   & $-7.94$ & $-10.73$  & $-3.40$ & $-6.18$ & $<-5.00$ & (5)\\
        \hline
        SN 2013am   &  0.9 & 20.9 & $4.1\pm1.3$  & $-6.67$ & $-7.11$  & $-2.98$ & $-3.58$\\
        SN 2013ej   &  2.2 & 12.5 & $2.5\pm0.8 $   & $-6.91$ & $-6.63$  & $-3.21$ & $-3.09$ & $-5.59$ & (6)\\
        SN 2016cok  &  0.2 & 59.0 &  $ 31.8\pm2.0  $ & $-5.13$ & $-4.83$  & $-1.90$ & $-1.79$ \\
        SN  2017eaw &  0.2 & 9.25 &  $ 2.1\pm0.1  $  & $-6.05$ & $-5.77$  & $-2.97$ & $-2.36$  & $-6.05$ & (7) \\
        \hline
    \end{tabular}
    \caption{The estimated ${}^{56}$Ni masses required for the observed late-time light curve to be powered entirely by radioactive decay. The $10^3L_\odot$, L${}_{SN}$, and $\beta_{SN}$ columns are the ${}^{56}$Ni mass corresponding to a luminosity of $10^3$ at time time $t_0$, the V band mean L${}_{SN}$ of the linear fit, and the slope $\beta_{SN}$ of the linear fit. References: (1) \citet{1992ApJ...398..248W}, (2) \citet{1996ApJ...461..993F}, (3) \citet{2006ApJ...641.1029C} , (4) \citet{2012ApJ...761..100C}, (5) \citet{2007ApJ...659.1488B}, (6) \citet{2016ApJ...817...22C}, (7) \citet{2018MNRAS.481.2536K} }
    \label{tab:analysis_results}
\end{table*}

\subsection{Radioactive Decay}

As discussed in \S~\ref{sec:radioactive}, we can estimate the required nickel mass by
scaling the V band light curve of SN~1987A either to the observed light
curve or the slope of the observed light curve.  To set the basic scale, Table~\ref{tab:analysis_results} gives the nickel mass corresponding to a V band luminosity of $10^3 L_\odot$ at time $t_0$. The SN~1987A V band light curve from \citet{2014ApJ...792...10S} extends to 4300 days.  If $t_0 < 4300$~days, we match the LBT and SN~1987A luminosities at $t_0$.  If $t_0>4300$~days but the time period used for the linear fit extends to times $<4300$~days (SN~2002hh, 2003gd), we use the linear fit estimate of the luminosity of the SN at $4300$~days.  If there is no overlap (SN 1980K, 1993J), we use the linear fit estimate for the earliest time included in the fits and the luminosity of 87A at 4300~days. This last case should underestimate the required mass. These mass estimates,
given in Table~\ref{tab:analysis_results}, are impossibly large.  Adding any additional luminosity (i.e. some multiple
of the mass corresponding to $10^3L_\odot$) will only exacerbate the problem.

We can avoid the problem of the reference image flux by instead fitting
the slope of the light curve, $\beta_{SN}$ at time $t_0$. These results are also given in  Table~\ref{tab:analysis_results}. For
systems where there is no overlap with the SN~1987A light curves, we use the slope at the end
of its light curve, which will again lead to an underestimate of the mass.  We again find unreasonably high $^{56}$Ni masses
for the SNe with well-measured slopes.  That late time emission cannot be powered by radioactivity is not surprising, but
this analysis emphasizes the degree to which it is infeasible.

\begin{table*}
    \centering
    \begin{tabular}{lcrrrrrrrr}
        \hline
           &   & & &\multicolumn{2}{c}{Mean Luminosity$[10^3L_\odot]$}    &   \multicolumn{3}{c}{Slope $[10^3L_\odot$yr${}^{-1}]$}   & \multicolumn{1}{c}{Ref Luminosity} \\
        SN & Band & $t_0$ & \multicolumn{1}{c}{L${}_{SN}$} & \multicolumn{1}{c}{$\langle$L${}_{i}\rangle$} & \multicolumn{1}{c}{$\sigma_{L_{i}}$} & \multicolumn{1}{c}{$\beta_{SN}$} & \multicolumn{1}{c}{$\langle\beta_i\rangle$} & \multicolumn{1}{c}{$\sigma_{\langle\beta_i\rangle}$} & \multicolumn{1}{c}{$[10^3L_\odot$]} \\
        \hline
        \hline

        SN 1980K  & U & 12414 & $-0.77\pm0.08$ & $6.48\pm0.46$ & $8.62$& $0.28\pm0.02$& $0.85$& $1.05$ & $0.4\pm0.2$ \\
                   & B & 12499 & $-1.35\pm0.09$ & $0.11\pm0.16$ & $0.97$& $-0.30\pm0.02$& $-0.01$& $0.08$ & $9.8\pm0.2$ \\
                   & V & 12602 & $-0.94\pm0.13$ & $0.01\pm0.07$ & $0.51$& $-0.07\pm0.04$& $0.01$& $0.05$ & $7.4\pm0.1$\\
                   & R & 12499 & $-1.52\pm0.08$ & $-0.07\pm0.09$ & $0.78$& $-0.29\pm0.02$& $0.02$& $0.08$ & $17.2\pm0.1$\\
          \\
        SN 1993J  & U & 7880 & $-2.59\pm0.10$ & $-0.03\pm0.08$ & $0.58$& $-0.68\pm0.03$& $0.02$& $0.14$ & $11.2\pm0.5$ \\
                   & B & 8983 & $-25.8\pm0.19$ & $-2.21\pm0.12$ & $7.55$& $-2.47\pm0.08$& $-0.37$& $0.67$ & $14.1\pm0.4$ \\
                   & V & 7630 & $-0.69\pm0.05$ & $-0.23\pm0.03$ & $0.69$& $-3.82\pm0.01$& $0.01$& $0.22$ & $12.0\pm0.2$\\
                   & R & 7839 & $-7.98\pm0.05$ & $-0.80\pm0.04$ & $2.38$& $-3.05\pm0.01$& $-0.30$& $0.86$ & $17.0\pm0.2$\\
          \\
        SN 2002hh & U & 4536 & $812\pm429$ & $-762\pm431$ & $1450$& $1680\pm118$& $3230$& $745$ & $50.2\pm6.3$   \\
                  & B & 4536 & $-337\pm66.9$ & $-181\pm67.4$ & $978$& $-200\pm18.0$& $-11.3$& $173$ & $52.5\pm7.0$\\
                  & V & 4536 & $-1470\pm14.7$ & $103\pm14.4$ & $186$& $-367\pm3.75$& $47.2$& $107$& $27.2\pm2.6$ \\
                  & R & 4536 & $-163\pm1.70$ & $-0.9\pm1.59$ & $23.3$& $-134\pm0.49$& $-1.80$& $7.47$ & $47.7\pm2.5$\\
          \\
        SN 2003gd & U & 4682 & $-1.42\pm0.28$ & $-0.46\pm0.29$ & $1.04$& $-0.31\pm0.07$& $-0.01$& $0.14$ & $110.0\pm2.4$\\
                  & B & 4387 & $-0.96\pm0.16$ & $-0.10\pm0.16$ & $0.68$& $-0.33\pm0.04$& $-0.01$& $0.18$ & $99.0\pm2.3$\\
                  & V & 4353 &$-1.16\pm0.11$ & $-0.09\pm0.11$ & $0.48$& $-0.75\pm0.03$& $0.01$& $0.17$ & $5.9\pm0.9$\\
                  & R & 4387 &$-1.10\pm0.08$ & $-0.04\pm0.08$ & $0.46$& $-0.20\pm0.02$& $-0.01$& $0.09$ & $6.4\pm0.8$ \\
          \\
        SN 2004dj & U & 4276 &$-2.31\pm0.09$ & $-0.09\pm0.05$ & $0.34$& $-1.20\pm0.03$& $0.03$& $0.03$ & $211.6\pm1.2$\\
                  & B & 3808 &$0.98\pm0.06$ & $-0.69\pm0.05$ & $1.37$& $-0.74\pm0.02$& $-0.05$& $0.15$ & $220.0\pm1.4$  \\
                  & V & 3808 & $1.82\pm0.04$ & $-0.32\pm0.03$ & $0.65$& $-0.60\pm0.01$& $-0.04$& $0.11$ & $111.9\pm1.3$  \\
                  & R & 3808 &  $2.54\pm0.05$ & $-0.02\pm0.04$ & $0.75$& $-0.29\pm0.01$& $0.02$& $0.12$ & $209.4\pm1.2$ \\
          \\
        SN 2005cs & U & 3509 &$-7.40\pm0.36$ & $1.91\pm0.33$ & $3.00$& $-1.73\pm0.14$& $0.01$& $0.57$ & $265.0\pm20.0$ \\
                  & B & 3347 & $2.21\pm0.27$ & $-0.31\pm0.25$ & $1.68$& $0.13\pm0.08$& $-0.07$& $0.57$ & $138.7\pm34.0$ \\
                  & V & 3677 & $-1.08\pm0.09$ & $-0.13\pm0.09$ & $0.59$& $-0.28\pm0.03$& $-0.08$& $0.24$& $72.9\pm7.8$ \\
                  & R & 3379 &$0.19\pm0.06$ & $-0.13\pm0.06$ & $0.34$& $0.01\pm0.02$& $0.01$& $0.13$ & $51.9\pm5.2$ \\
          \\
        \hline
    \end{tabular}
    \caption{Linear fits $L =L_{SN} + \beta_{SN} (t-t_0)$ to the late time light curves with $t_0$ being the mean time of the fitted points in days after peak. The mean luminosity, slope, and dispersions of the comparison light curves are given by $\langle L_{i} \rangle$, $\langle \beta_i \rangle$, $\sigma_{L_{i}}$, and $\sigma_{\langle\beta_i\rangle}$.}
    \label{tab:SN Decay}
\end{table*}

\begin{table*}
    \centering
    \begin{tabular}{lcrrrrrrrr}
        \hline
           &   & & &\multicolumn{2}{c}{Mean Luminosity$[10^3L_\odot]$}    &   \multicolumn{3}{c}{Slope $[10^3L_\odot$yr${}^{-1}]$}   & \multicolumn{1}{c}{Ref Luminosity} \\
        SN & Band & $t_0$ & \multicolumn{1}{c}{L${}_{SN}$} & \multicolumn{1}{c}{$\langle$L${}_{i}\rangle$} & \multicolumn{1}{c}{$\sigma_{L_{i}}$} & \multicolumn{1}{c}{$\beta_{SN}$} & \multicolumn{1}{c}{$\langle\beta_i\rangle$} & \multicolumn{1}{c}{$\sigma_{\langle\beta_i\rangle}$} & \multicolumn{1}{c}{$[10^3L_\odot$]} \\
        \hline
        \hline
        SN 2009hd & U & 2057 & $1620\pm67.5$ & $-202\pm63.9$ & $452$& $-134\pm28.0$& $-32.1$& $110$ & $133\pm14.2$ \\
                  & B & 2417 &  $1260\pm30.2$ & $-68.8\pm28.7$ & $311$& $-239\pm9.59$& $-9.19$& $42.7$ & $267\pm27.8$\\
                  & V & 2417 & $839\pm6.17$ & $17.6\pm6.16$ & $123$& $-63.5\pm1.94$& $5.97$& $16.6$ & $114\pm12.3$\\
                  & R & 2608 & $995\pm7.12$ & $-323\pm7.20$ & $937$& $-94.6\pm2.31$& $3.98$& $14.0$ & $164\pm29.3$ \\
          \\
        SN 2011dh & U & 2455 & $-20.8\pm0.49$ & $0.48\pm0.46$ & $4.16$& $-0.43\pm0.19$& $0.24$& $0.42$ & $28.3\pm2.6$\\
                  & B & 2171 & $-88.5\pm0.25$ & $-0.45\pm0.24$ & $1.97$& $-1.38\pm0.09$& $0.06$& $0.43$ & $50.2\pm3.0$\\
                  & V & 2158 &$-86.8\pm0.14$ & $-0.11\pm0.09$ & $1.14$& $-1.64\pm0.06$& $0.07$& $0.09$ & $38.3\pm1.7$\\
                  & R & 1907 & $-97.3\pm0.16$ & $-1.00\pm0.16$ & $2.34$& $-1.38\pm0.09$& $0.02$& $0.30$ & $60.5\pm1.9$\\
          \\
        SN 2012fh & U & 1932 & $-10.05\pm0.28$ & $-0.04\pm0.23$ & $1.87$& $-0.61\pm0.11$& $-0.08$& $0.28$ & $45.1\pm3.9$ \\
                  & B & 1932 & $-2.50\pm0.19$ & $-0.53\pm0.17$ & $2.34$& $-0.17\pm0.07$& $-0.22$& $0.30$ & $8.80\pm2.7$ \\
                  & V & 1919 & $-0.25\pm0.11$ & $0.31\pm0.10$ & $0.51$& $0.02\pm0.05$& $-0.04$& $0.18$ & $3.90\pm0.8$ \\
                  & R & 1932 & $1.17\pm0.17$ & $1.32\pm0.16$ & $1.50$& $0.64\pm0.07$& $-0.05$& $0.33$ & $27.6\pm1.5$\\
          \\
        SN 2013am & U & 1300 & $34.96\pm6.60$ & $3.75\pm6.72$ & $59.71$& $-5.78\pm3.48$& $-0.91$& $15.45$ & $4.0\pm1.5$   \\
                  & B & 1795 & $14.66\pm3.05$ & $5.12\pm3.11$ & $13.02$& $-4.07\pm1.34$& $1.10$& $3.23$ & $7.8\pm2.5$ \\
                  & V & 1840 & $16.90\pm1.12$ & $0.12\pm1.13$ & $6.69$& $-2.39\pm0.53$& $-0.23$& $1.30$ & $8.8\pm2.5$ \\
                  & R & 1853 & $6.96\pm1.01$ & $-1.10\pm1.01$ & $6.66$& $-0.41\pm0.48$& $0.25$& $1.53$ & $19.1\pm4.2$ \\
         \\          
        SN 2013ej & U & 1915 & $9.89\pm0.30$ & $0.09\pm0.37$ & $1.40$& $-1.54\pm0.15$& $0.02$& $0.35$ & $14.2\pm1.1$  \\
                  & B & 1941 & $15.91\pm0.21$ & $-0.90\pm0.40$ & $1.79$& $-1.86\pm0.10$& $-0.29$& $0.44$ & $11.8\pm1.3$\\
                  & V & 2267 & $13.26\pm0.23$ & $-0.11\pm0.18$ & $1.55$& $-1.45\pm0.16$& $0.02$& $0.23$ & $10.6\pm0.5$\\
                  & R & 1915 & $10.46\pm0.16$ & $-0.76\pm0.17$ & $1.49$& $-1.22\pm0.08$& $-0.03$& $0.17$ & $9.5\pm0.8$ \\
          \\
        SN 2016cok & U & 1578 & $61.60\pm4.94$ & $1.86\pm4.97$ & $14.8$& $-4.09\pm9.83$& $7.41$& $37.65$& $29.1\pm3.6$ \\
                   & B & 1410 & $396.37\pm5.59$ & $1.36\pm5.45$ & $37.09$& $-56.35\pm6.17$& $14.07$& $28.74$ & $60.3\pm8.2$\\
                   & V & 1396 & $201.77\pm1.68$ & $3.61\pm1.62$ & $12.43$& $-118.29\pm1.82$& $-2.78$& $7.29$ & $26.5\pm6.4$ \\
                   & R & 1410 & $390.02\pm2.83$ & $8.74\pm2.75$ & $26.19$& $-78.02\pm3.25$& $-1.36$& $10.02$ & $59.4\pm8.6$\\
          \\
        SN 2017eaw & U & 1245 &$77.23\pm1.23$ & $-1.35\pm1.21$ & $3.15$& $-14.16\pm1.35$& $0.16$& $1.91$& -- \\
                   & B & 1261 &$94.43\pm0.66$ & $1.11\pm0.63$ & $2.92$& $-18.05\pm0.79$& $1.31$& $4.39$ & $1.3\pm0.2$\\
                   & V & 1245 & $60.55\pm0.33$ & $0.51\pm0.33$ & $3.28$& $-7.72\pm0.36$& $-0.01$& $0.82$& $1.0\pm0.1$ \\
                   & R & 1261 & $48.20\pm0.25$ & $0.90\pm0.24$ & $2.01$& $-8.97\pm0.32$& $0.27$& $1.35$& $0.9\pm0.1$\\
        \hline
        
    \end{tabular}
    \caption{Linear Fits Continued.}
    \label{tab:SN Decay Cont.}
\end{table*}

\subsection{CSM Interactions}

\begin{figure}
    \centering
    \includegraphics[width=\columnwidth]{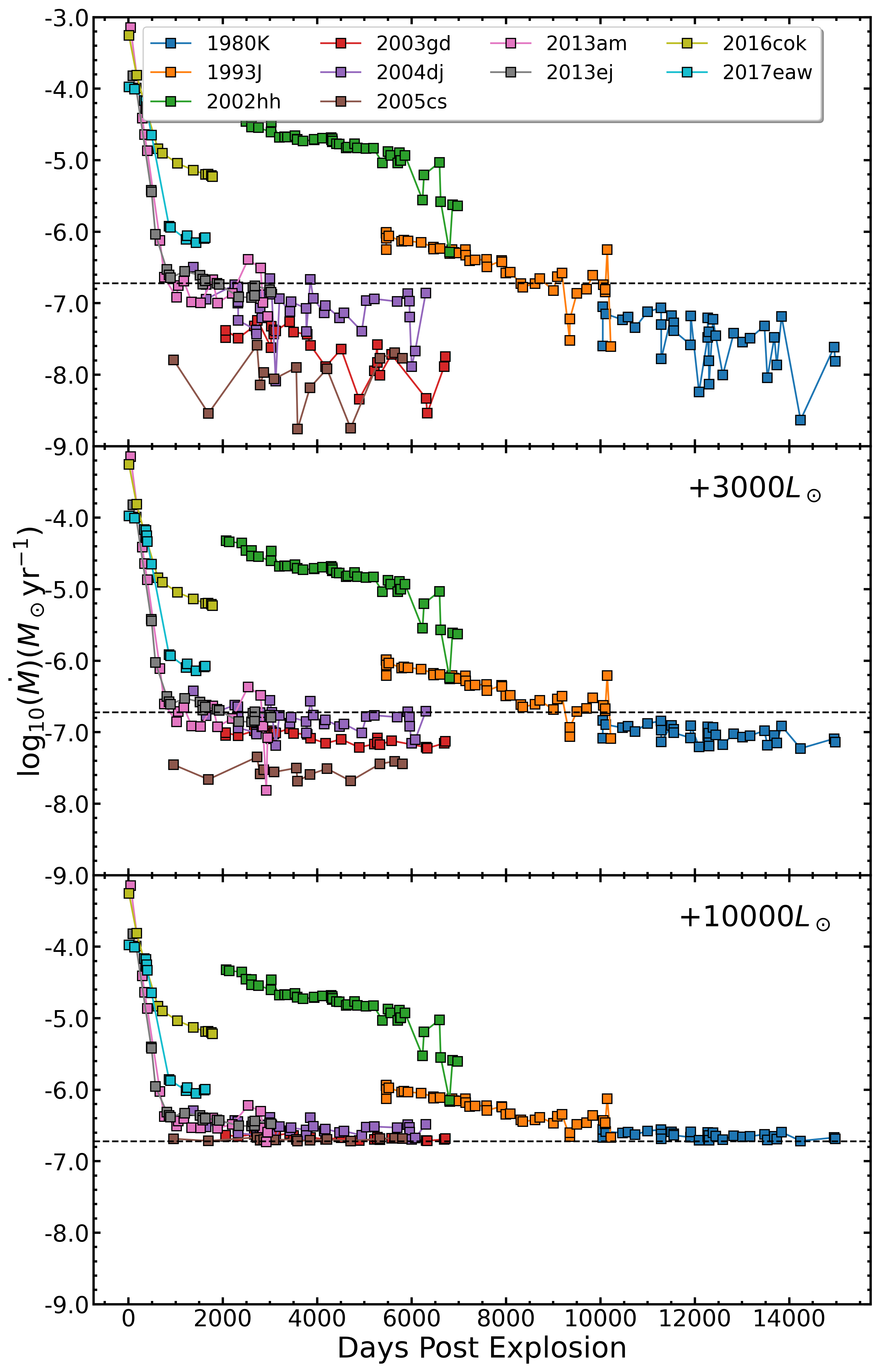}
    \caption{Pre-SN mass loss rates required to produce the observed R band luminosity after scaling $\nu L_\nu$ after adding zero (top), 3000L$_\odot$ (middle), and 10000L$_\odot$ (bottom) for the flux in the reference image.
    The dashed line represents a $\dot{M}$ corresponding to a luminosity of $10^4$L$_\odot$.} 
    \label{fig:all_CSM}
\end{figure}

Table~\ref{tab:analysis_results} gives two estimates of $\dot{M}$, both under the assumption that $v_s=4000$~km/s, $v_w=10$~km/s and $\epsilon=0.1$ (Eqn.~\ref{eqn:CSM}). The first is based on the luminosity at time $t_0$ after shifting $L_{SN}$ upwards to make the linearly interpolated luminosity at the last epoch to be zero if it is negative.  
 The second is simply to use the change in luminosity $\beta_{SN} \Delta t$ over $\Delta t=10$~years.  We can also visualize the required mass loss rates as shown in Fig.~\ref{fig:all_CSM}.   The resulting estimates generally range from
$10^{-7} M_\odot$/year to $10^{-5} M_\odot$/year. Adding a luminosity
of $10^3$, $3\times 10^3$, $10^4$ or $3\times 10^4 L_\odot$ for the flux in the reference image,
corresponds to increasing the mass loss rates by
$\dot{M} = 10^{-7.72}$, $10^{-7.24}$, $10^{-6.72}$ and $10^{-6.24} M_\odot$/year.  We illustrate
the cases of adding $3\times 10^3$ and $10^4 L_\odot$ in Fig~\ref{fig:all_CSM}. If we increase the reference
frame flux much more, we would start trivially seeing a source -- for example, SN~1980K is a clear source in the
R band reference image with a luminosity of $1.7 \times 10^4 L_\odot$, quite consistent with the $1.3 \times 10^4 L_\odot$ found with HST just before the start of the LBT observations (see Table~\ref{tab:HST_DATA}).  Adding this to our
standard estimate just using the difference light curve only increases $\dot{M}$ from $10^{-6.65}$ to $10^{-6.26}M_\odot$/year - a quantitatively important change but one without any importance for our qualitative conclusions.
Table~\ref{tab:analysis_results} also
reports the mass loss rate corresponding to the estimated drop in luminosity $\beta_{SN} \Delta t$ over $\Delta t=10$~yr,
which gives similar results.  
There is considerable freedom in the absolute scales from choosing the parameters, particularly the shock velocity $v_s$ and the radiative efficiency $\epsilon$ in Eqn.~\ref{eqn:CSM}.

Figure~\ref{fig:CSM_color_comp} shows the V$-$R and B$-$V colors of SN~1987A \citep{2019ApJ...886..147L} and the model spectrum of \citet{2022A&A...660L...9D}, which is a good fit to late time
spectra of SN~1993J.  We chose these colors because the B and R bands contain strong hydrogen Balmer emission lines while the V band only has weaker lines in the \citet{2022A&A...660L...9D} model spectra.  These leads to red V$-$R colors and blue B$-$V colors that are difficult for a continuum emission source to mimic, as illustrated by the colors of the PARSEC stellar isochrones. 

The inferred mass loss rates are modest, ranging from $\sim 10^{-7}$ to $\sim 10^{-5} M_\odot\hbox{yr}^{-1}$.  For comparison, Table~\ref{tab:analysis_results} gives
mass loss estimates for these SNe from the literature.  These are generally post-explosion estimates based on radio and
X-ray observations, although there are exceptions (see the Appendix).  None of the estimates are based on the late time
optical emission, although late time CSM emission is well established for SN~1980K and 1993J \citep{2012ApJ...751...25M}
(as had already been noted by \citealt{2021MNRAS.508..516N}), and SN~2004dj (\citealt{2012ApJ...761..100C}, \citealt{2018ApJ...863..163N}).  Recently, \citet{2022MNRAS.tmp.3313V} also found that the progenitors of SN~2013ej and SN~2017eaw are still brighter than their progenitors
in the V band, but not in the I band, in 2021 and 2020, respectively, and hypothesize that this may be due to ongoing CSM interactions.  In both the \cite{Matheson2000} spectra of SN~1993J and the theoretical spectra of \citet{2022A&A...660L...9D} 
there are strong lines in the UBVR bands but not in the I band, consistent with these observations.

\subsection{Dust Echoes}

\begin{figure}
    \centering
    \includegraphics[width=\columnwidth]{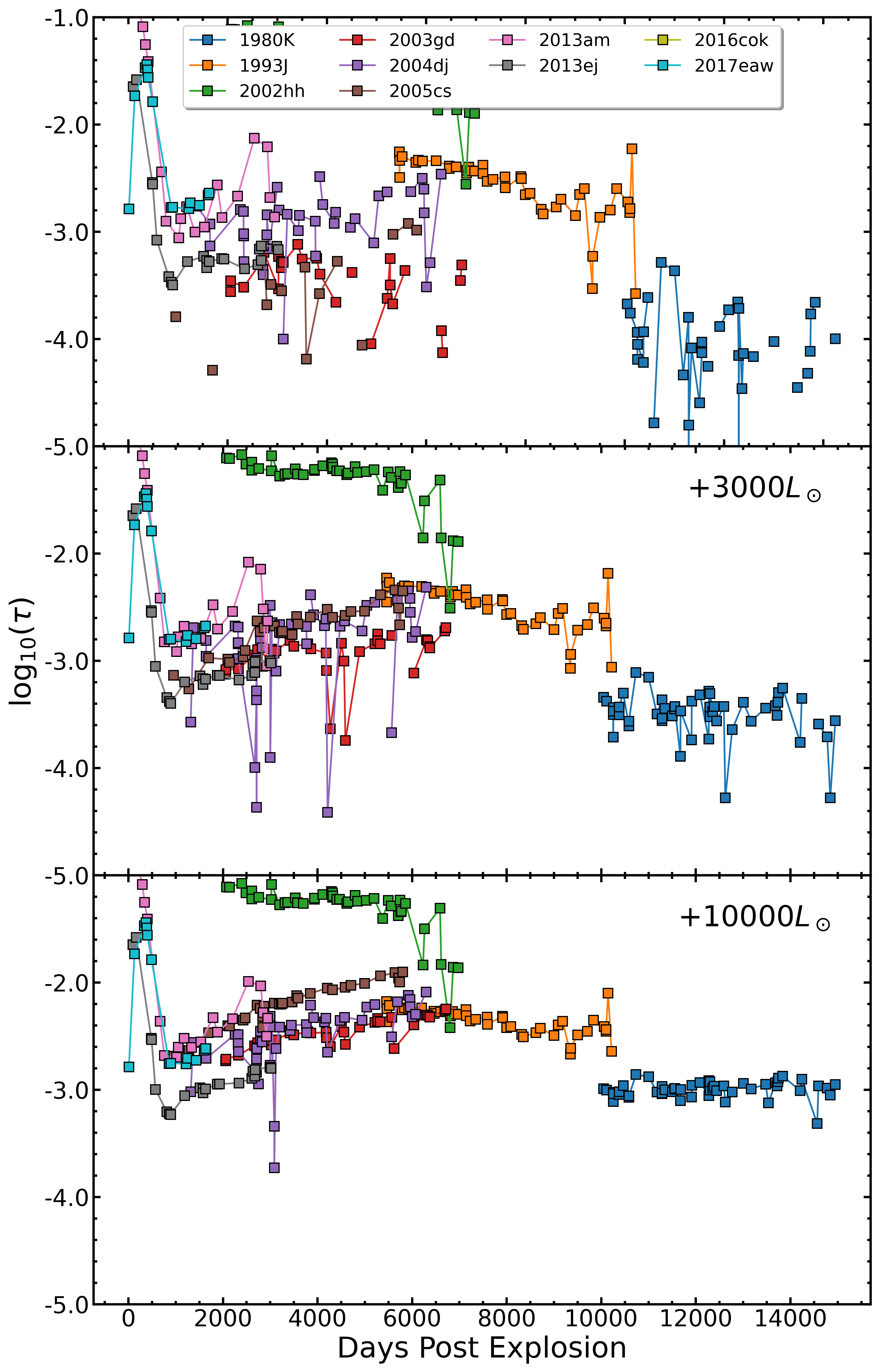}
    \caption{Required scattering optical depth $\tau$ (Eqn. \ref{eqn:echo}) for the observed R band luminosity (V band for SN~1980K) after adding zero (top), 3000L$_\odot$ (middle), and 10000L$_\odot$ (bottom) for the flux in the
     reference image.}
    \label{fig:lightecho}
\end{figure}

Using the same procedures as in \S3.3 and  Eqn.~\ref{eqn:echo} yields the optical depth estimates given in
Table~\ref{tab:analysis_results}.  The required optical depths are not very large.  We can recast Eqn.~\ref{eqn:echo}
using $E_{rad} = L_{peak} t_{peak}$ where $L_{peak} \sim 10^8 L_\odot$ is the peak luminosity and $t_{peak} \simeq 100$~days
is the duration of the plateau phase, so that $ \tau \sim (L_{SN}/L_{peak})(t_0/t_{peak})$.  The luminosity has dropped
by $L_{SN}/L_{peak} \sim 10^4$ or more, while the elapsed time is only $t_0 \sim 30 t_{peak}$, leading to required optical
depths of $\tau \sim 10^{-3.5}$-$10^{-2.5}$.  However, as illustrated in Fig.~\ref{fig:lightecho}, the shapes of the light
curves generally show phases with the optical depth increasing with time.
We also see in Figure~\ref{fig:CSM_color_comp} that the colors of the late time emission are different
from the expectations for dust echoes.  The dust echo colors should, essentially, be the colors of the SN shifted
roughly parallel to the isochrones while the observed colors are generally shifted perpendicular to the isochrones.
These facts appear to rule out dust echoes as a general explanation of the late time emission.

At first glance, these low required optical depths might seem to require significant dust echoes from
essentially all the SNe because the required optical depths are much lower than the extinction estimates in 
Table~\ref{tab:SN_INFO} since $\tau_R \simeq 2.3 E(B-V)$.  The missing element is that 
the dust echo at any given time is being produced by a very thin region of thickness 
$\sim c t_{peak} \sim 0.1$~pc corresponding to the spatial extent of the luminosity transient.
If the dust is uniformly distributed over a line of sight distance of 0.1 to 1~kpc, the optical
depth scale relevant to producing an echo is fraction $10^{-3}$ to $10^{-4}$ of the total along 
the line of sight.  Concentrating the dust in ``sheets'' helps, but more at early times when the
physical region producing the echo is small.  This discussion assumes unresolved echoes where the
surface brightness of the echo is not relevant to its detection. 

The exception is SN~2002hh which is both heavilty obscured (Table~\ref{tab:SN_INFO}) and known to have a 
strong dust echo (e.g., \citealt{2005ApJ...627L.113B}, \citealt{2006MNRAS.368.1169P}, \citealt{2007ApJ...669..525W}, and \citealt{2012ApJ...744...26O}).  With the large corrections for extinction, SN~2002hh has the greatest late-time
luminosity of any of the 12 systems by almost two orders of magnitude.  While it is fading in the R band,
and evolving little in the B and V bands, the U band luminosity is clearly increasing.  It is the only such
system.

\subsection{Binary Shock Interaction}

\begin{figure}
    \centering
    \includegraphics[width=\columnwidth]{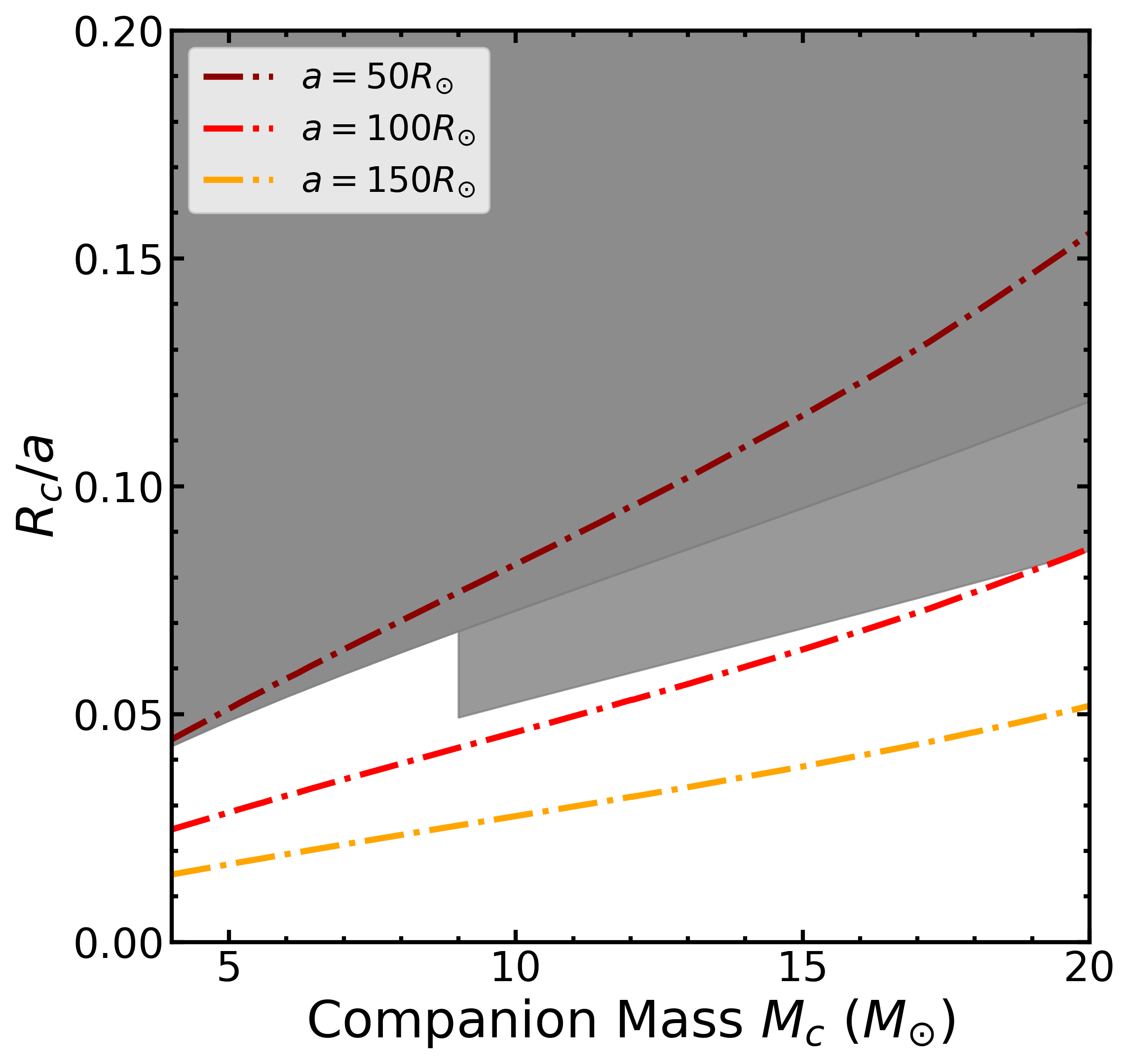}
    \caption{Allowed binary companions for SN~2012fh. The gray region excludes companions on their predicted inflation timescales and maximum luminosity. The brown, red, and orange dot-dashed lines represent orbital separations of 50, 100, and 150 R$_\odot$ respectively, with stellar counterpart masses and radii obtained from a solar metalicity, PARSEC isochrone with an age of $10^{6.7}$ years. This age is selected so that M$<20M_\odot$ have not yet evolved off the main sequence.}
    \label{fig:12fh_ratio_mass}
\end{figure}

We only consider the Type~Ib/c SN~2012fh for this case because the progenitors of all the other SNe should be giants where it is impossible to significantly heat a main sequence companion as they subtend such a small solid angle. 
The companion star to a stripped SN should generally be a cooler
star and dominate the optical emission even before being heated, so we assume that we can neglect the effect of the progenitor
luminosity on the subtracted light curves.  Under these assumptions, we estimate that a companion can have increased in 
luminosity by no more than $L_{max}\simeq10^5L_\odot$ (Eqn.~\ref{eqn:L_max}).
Based on Fig.~\ref{fig:MultiStackIb/c}, SN~2012fh faded within ${\sim}$440 days after peak luminosity, which we use 
as our limit on the inflation time scale $\tau_{infl}$ (Eqn.~\ref{eqn:Infl_T}). Fig.~\ref{fig:12fh_ratio_mass} translates these limits
into the allowed companion mass $M_c$ and radius relative to the binary semi-major axis $R_c/a$.

Using Solar metallicity \verb"PARSEC" isochrone with $\log_{10}$(Age) = 6.7, for which no $M_c<20M_\odot$ companions have evolved, we also show curves with constant semi-major axes of $a=50 R_\odot$, $100R_\odot$ and $150 R_\odot$ given the stellar radii $R_c$ of the models.  At least for these models, the binary would have to be wider separation than $50$-$100R_\odot$ at the time of the explosion. If the progenitor of SN~2012fh was stripped through binary interactions, then the mass-radius relations of the single-star \verb"PARSEC" isochrones will be incorrect at some level.

\section{Conclusions} \label{sec:results}

Motivated by the discovery that SN~2013am and SN~2013ej were still optically brighter than their progenitors a decade after explosion
(\citealt{2021MNRAS.508..516N}), we systematically investigated the evolution over the last 14 years of the
12 ccSNe that occurred in the LBT 
search for failed SNe (\citealt{2008ApJ...684.1336K}, \citealt{Gerke2015}, \citealt{2017MNRAS.468.4968A}, \citealt{2017MNRAS.469.1445A}, \citealt{2021MNRAS.508.1156B}, \citealt{2021MNRAS.508..516N}) galaxies from 1980 onwards.
We used difference imaging techniques to look for 
continued evolution in their UBVR luminosities.  Difference imaging has the advantage of largely eliminating the problem of crowding
in these ground based data, but some quantitative conclusions depend on the difficult to estimate flux of the targets
in the reference image.  As part of the survey, we analyzed the SED of the progenitor of SN~2011dh
from the LBT data and find results consistent with the results from pre-SN HST data (\citealt{2011ApJ...741L..28V}, \citealt{2011ApJ...739L..37M}).  This analysis also demonstrates that the LBT easily has the sensitivity to probe
emissions at the level of $\sim 10\%$ of the progenitor luminosity or better (see Fig.~\ref{fig:11dhsed}). We also considered shock heated binary companions to the Type~Ibc SN~2012fh and set limits on the companion
mass $M_c$ and the ratio of the companion radius to the semi-major axis $R_c/a$ (see Fig.~\ref{fig:12fh_ratio_mass}).

We focus these conclusions on the interesting finding that of the 11 Type~II SNe in the sample, only two
(the Type~IIP SN~2005cs and the Type IIb SN~2011dh), do not show continued, evolving SN emission 5-42 years
post-explosion.  This includes the two oldest systems studied, the Type~IIL SN~1980K and the Type~IIb SN~1993J.
The continued emission is coincidentally on the scale of the luminosity of the progenitors 
($\sim 10^4 L_\odot$). For the 6
of these SNe that occured after the LBT survey started, we know this is true of five (the exception is SN~2011dh)
because we constructed the reference image from pre-SN images so that the luminosities in the subtracted light
curves are relative to the luminosity of the progenitor.  Apparently, SN like SN~1987A (e.g., \citealt{1988ApJ...330..218W}, \citealt{2007Msngr.127...44F}, \citealt{2014ApJ...792...10S}) or SN~2011dh which 
rapidly fade to luminosities well below those of their progenitors are the exceptions, not the norm.

For completeness, we considered radioactivity, CSM interactions, dust echoes and magnetar/engines as possible
drivers.  Radioactivity requires impossible masses and dust echoes generally require too much dust. Neutron star spin
down can produce the necessary luminosity, but is not believed to produce significant optical emission at these
phases (e.g., \citealt{2010ApJ...717..245K}), which is certainly true of the Crab pulsar today.  The time 
evolution of the luminosity and
the colors of the emission are also generally inconsistent with dust echoes.   Not surprisingly, since CSM emission
has been observed at earlier phases for many of these sources (see Table~\ref{tab:analysis_results} and Appendix~\ref{appendix}), the only logical 
possibility is continuing  CSM emission aside from SN~2022hh where it is probably due to a very strong,
continuing dust echo.

Mass loss is a crucial component of RSG evolution (e.g., \citealt{1975MSRSL...8..369R}, \citealt{1988A&AS...72..259D}, \citealt{1990A&A...231..134N}, \citealt{2005A&A...438..273V}, \citealt{2011A&A...526A.156M}, \citealt{2021ApJ...922...55B}).  If we use the progenitor luminosity scaling from Eqn.~\ref{eqn:lprog}, we can
express the \citet{1990A&A...231..134N} mass loss rate as
\begin{equation}
   \dot{M} = 10^{-5.60}\left( \frac{ M}{10 M_\odot} \right)^{2.62} \left( \frac{T_e}{3500\hbox{K}} \right)^{-1.61}M_\odot~\hbox{yr}^{-1}
\end{equation}
and the \citet{2005A&A...438..273V} mass loss rate as
\begin{equation}
   \dot{M} = 10^{-4.81}\left( \frac{ M}{10 M_\odot} \right)^{1.58} \left( \frac{T_e}{3500\hbox{K}} \right)^{-6.3}M_\odot~\hbox{yr}^{-1}.
\end{equation}
With the temperature fixed at $3500$~K, the mass loss rates for $8M_\odot < M < 20M_\odot$ span $-5.9 < \log\dot{M} < -4.8$
and $-5.6 < \log\dot{M} < -4.8$.  These estimates are somewhat higher than our estimates in Table~\ref{tab:analysis_results}. While there are significant quantitative uncertainties in these estimates, particularly through the assumed shock velocity ($v_s=4000$~km/s) and the
radiative efficiency ($\epsilon=0.1$) it would be difficult to significantly raise them given the
lack of large numbers of X-ray bright Type~II SNe (see \citealt{2014MNRAS.440.1917D}).
There are also arguments that these models for RSG mass loss rates are too
high (e.g., \citealt{2021ApJ...922...55B}). 

Nonetheless, if we use these mass loss rates to predict the CSM luminosity (Eqn.~\ref{eqn:CSM}) relative
to the progenitor luminosity (Eqn.~\ref{eqn:lprog}), we find that
\begin{multline}
    \frac{ L_{shock}}{L_{prog}} = 2.09 \left(\frac{\epsilon}{0.1}\right) \left(\frac{ M }{10M_\odot}\right)^{1.12}
         \left( \frac{ T }{3500~\hbox{K}}\right)^{-1.61} \\
          \times \left( \frac{ v_s}{4000~\hbox{km/s}}\right)^3
          \left( \frac{ 10~\hbox{km/s}}{v_w }\right)
\end{multline}
and
\begin{multline}
    \frac{ L_{shock}}{L_{prog}} = 12.9 \left(\frac{\epsilon}{0.1}\right) \left(\frac{ M }{10M_\odot}\right)^{0.08}
         \left( \frac{ T }{3500~\hbox{K}}\right)^{-6.3}  \\
         \times \left( \frac{ v_s}{4000~\hbox{km/s}}\right)^3
         \left( \frac{ 10~\hbox{km/s}}{v_w }\right)
\end{multline}
for the two mass loss prescriptions.  Since the luminosity of an RSG progenitor in a particular band is less
than $L_{prog}$, it is clear that CSM emission from normal RSG winds can easily exceed the band luminosity of the progenitor provided the efficiency of converting the shock luminosity into optical emission is $\epsilon \gtorder 0.01$.  
The interesting question then becomes the duration of the emission, since we generically find that the emissions are fading. One possibility, requiring theoretical study, is the physics of the efficiency $\epsilon$. While the shock luminosity for a fixed shock velocity is constant for a $1/r^2$ wind density, it would still be 
fairly natural for $\epsilon$ to evolve since most emission mechanisms scale as the square of the density
rather than being proportional to the density like the shock luminosity. 
The fading could also be due to having wind density profiles that are generically steeper than the $1/r^2$ density
profile of a constant $\dot{M}$ wind (see, e.g., \citealt{2012MNRAS.419.1515D}).

The missing element for making these conclusions more quantitative is to obtain absolute calibrations.  Waiting
for the SN to fade to black seems to require challenging levels of patience given that SN~1980K seems to still 
be going strong. On the other hand, SN~1993J does appear to be close to fading away.   Fortunately, single orbit
HST observations can provide the necessary absolute flux levels for two to three filters depending on the
desired level of control over systematic problems (cosmic rays, bad pixels etc.).  For two filters, the preferred
choices are probably R and B given the properties of observed (\citealt{Matheson2000}) and theoretical (\citealt{2022A&A...660L...9D}) spectra of CSM interactions, with adding V if a third filter is included.

The SN shock also eventually reaches the edge of the wind. In the \citet{1976ApJS...32..233W} self-similar solution for a wind
expanding into a constant density medium, the contact discontinuity between the two fluids lies at
\begin{multline}
   R_{CD} = 5.4 \left( \frac{ \dot{M}}{10^{-6}M_\odot\hbox{yr}^{-1}} \right)^{1/5} 
         \left( \frac{ v_w}{10~\hbox{km/s}} \right)^{2/5} \\
         \times \left( \frac{ t}{10^6~\hbox{yr}}\right)^{3/5}
             \left( \frac{ \hbox{cm}^{-3}}{n }\right)^{1/5}~\hbox{pc}
\end{multline}
where $n$ is the ambient density and $t$ is the time since the start of the wind.  Except for time, the
parameter dependence is weak, so the time scale for the SN shock to reach the contact discontinuity is
in unfortunately long ($1300$~years for $v_s=4000$~km/s).  Moreover, the RSG wind may be expanding into
a lower density bubble formed by a faster main sequence wind, leading to a still more distant contact
discontinuity (e.g., \citealt{2005ApJ...630..892D}).  The wind termination shock can
lie at a significantly smaller radius than the contact discontinuity, but centuries are not much of
a gain over millennia.  

Even if it seems unlikely that we will observe these boundaries of the RSG wind, it is still of interest
to both continue to observe the evolution of these SNe and to extend the study to the small numbers of 
still older historical SNe in the LBT galaxies.  Moreover, the estimated time scale to reach the edge 
of the wind is just an estimate -- in SN~1987A, the shock started to interact with the wind boundary 
after only 20 years (e.g.,  \citealt{2019ApJ...886..147L}).  SN~2002hh is postulated to have a massive
($\sim 0.1M_\odot$) dusty shell at a distance of $0.03$ to $0.3$~pc (e.g., \citealt{2005ApJ...627L.113B}).
Like \citep{2015MNRAS.451.1413A}, we see no evidence for interactions with the postulatedshell, which leads to a minimum distance to the shell of $0.08(v_s/4000~\hbox{km/s})$~pc that begins
to strongly constrain these models.
And, since they are all fading, it will eventually be
possible to do the progenitor photometry of those SNe with pre-explosion LBT imaging - just after far
more time than expected.

\section*{Acknowledgements}

We thank Krzysztof Stanek for his assistance with LBT calibrations and troubleshooting. We also thank Todd Thompson for valuable discussions. J.N. and C.S.K are supported by NSF grants AST1814440 and AST-1908570. The LBT is an international collaboration among institutions in the United States, Italy and Germany. LBT Corporation partners are: The University of Arizona on behalf of the Arizona university system; Istituto Nazionale di Astrofisica, Italy; LBT Beteiligungsgesellschaft, Germany, representing the Max-Planck Society, the Astrophysical Institute Potsdam, and Heidelberg University; The Ohio State University, and The Research Corporation, on behalf of The University of Notre Dame, University of Minnesota and University of Virginia.

\section*{Data Availability}

The data underlying this article will be shared on reasonable
request to the corresponding author.



\bibliographystyle{mnras}

\bibliography{refs} 




\appendix



\section{Individual Supernovae}\label{appendix}
Here we summarize the individual SNe. Table~\ref{tab:SN_INFO} lists the SN and gives the adopted distances, Galactic and host extinctions. Table~\ref{tab:HST_DATA} lists HST observations that can be used to provide absolute calibrations of the LBT light curves, and Table~\ref{tab:analysis_results} gives previous estimates of mass loss rates $\dot{M}$.

\subsection*{SN 1980K}

The Type II-L SN 1980K was discovered in NGC 6946 \citep{SN1980K} on 1980 October 28 reaching a maximum of $V=11.4$ mag on November 3 1980 \citep{Barbon1980} and has been extensively studied far into the nebular phase (e.g., \citealt{Chevalier1982} \citealt{1983ApJ...274..175D}, \citealt{1998ApJ...506..874M}, \citealt{1990ApJ...351..437F}, \citealt{1991ApJ...372..531L}, \citealt{1994ApJ...420..268C}, \citealt{2012ApJ...751...25M}). Based on the radio emission, \cite{Chevalier1982} estimated a rough
mass loss rate of $\dot{M} \sim 10^{-5}M_\odot$~year$^{-1}$ for $v_w = 10$~km/s, and \cite{1998ApJ...506..874M} argue for a break to a region with a lower wind density due to a sudden drop in the radio fluxes around 1995.  \cite{1983ApJ...274..175D} argued for a dust echo from a shell formed at the wind/ISM boundary with a detailed model in \cite{2012ApJ...749..170S} placing the shell
$\sim 15$~pc from the SN.  
They find weak evidence for optical fading between 2005 and 2008 and strong evidence for
an infrared excess. The source they identify as the SN is clearly visible in the first LBT images, taken 4 months after the
\citet{2012ApJ...749..170S} HST observations in January 2008. 
The LBT observations of NGC~6946 now span 13 years with 
approximately 57 epochs per filter.

\subsection*{SN 1993J}
SN 1993J in M81 is the prototype of the Type IIb SNe class (\citealt{1993ApJ...419L..73S}, \citealt{1993Natur.364..600S}, \citealt{1993ApJ...415L.103F}, \citealt{1993Natur.364..507N}, \citealt{1993JIIb}). It was discovered on 1993 March 28 \citep{1993IAUC.5731....1R}. Using ground based observations obtained prior to the SN, the progenitor of 1993J was identified as K0 RSG \citep{1993JProgenitor}, and its eventual disappearance has been observed \citep{2009Sci...324..486M}.   The progenitor was not variable by more than 0.2~mag (\citealt{1995ApJ...455..658S}).
There is extensive evidence for CSM emission from late time spectra
(e.g., \citealt{2000AJ....120.1487M}, \citealt{2000AJ....120.1499M}), X-ray emission (e.g., \citealt{1993ApJ...419L..73S}, \citealt{1994ApJ...431L..95L}, \citealt{1995ApJ...455..658S}, \citealt{1996ApJ...461..993F}, \citealt{2019ApJ...875...17K}), and radio
emission (e.g., \citet{1994ApJ...432L.115V}, \citealt{2019ApJ...875...17K}).  VLBI observations monitored the expansion of the 
remnant for almost a decade \citep{2003ApJ...597..374B}, tracking the expansion to 19,000~AU (0.1~pc) and implying
average expansion velocities of $\sim 10^4$~km/s.  The \citet{2019ApJ...875...17K} models assume a mass loss rate of 
$\dot{M}=4 \times 10^{-5} M_\odot$/year before $3000$~days. The LBT observations of M81 now span ${\sim}$14 years with approximately 32 epochs per filter.  For SN~1993J, the HST observations of \citet{2014ApJ...790...17F} provide an
absolute calibration of the V and R band light curves (Table~\ref{tab:HST_DATA}).

\subsection*{SN 2002hh}

SN 2002hh is a Type IIP SN first discovered on 2002 October 31.1 \citep{2002IAUC.8005....1L} by LOSS in NGC 6946. 
\citet{2005ApJ...627L.113B} argue for a massive ($\sim 0.1M_\odot$) dusty shell at a distance of $\sim 0.04$~pc,
and the dust echoes are discussed further in \cite{2006MNRAS.368.1169P}, \cite{2007ApJ...669..525W}, and \cite{2012ApJ...744...26O}.  \citet{2015MNRAS.451.1413A} spectroscopically searched for any evidence
of shock interactions with such a dusty shell in 2013, 10.5 years after explosion, finding none. The
spectra appear similar to the SN near peak, suggesting that the emission is dominated by a dust
echo.  We can use the HST
observations from \cite{2012ApJ...744...26O} to normalize some of the LBT bands (Table~\ref{tab:HST_DATA}).  
This galaxy has been observed by the LBT since mid-2008 with a total of 57 epochs in each filter.

\subsection*{SN 2003gd}

SN 2003gd is a Type IIP SN discovered in NGC 628 (M 74) on 2003 June 12.82 \citep{2003IAUC.8150....2E} with
a general study in \cite{2005MNRAS.359..906H}.  Archival Hubble Space Telescope (HST) and Gemini data identified its progenitor, a RSG star with an estimated mass of $ 8^{+4}_{-2}M_\odot$ (\citealt{2004Sci...303..499S}, \citealt{2005MNRAS.359..906H}) and the progenitor is observed to have disappeared (\citealt{2009Sci...324..486M}, \citealt{2014MNRAS.438..938M}). Two years after explosion it had a faint, resolved (by HST) dust echo with 
$\sim 1\%$ of the optical flux of the SN \citep{2005ApJ...632L..17S}.  
This galaxy has been observed for a total of 32 epochs. 

\subsection*{SN 2004dj}
SThe Type~IIP SN 2004dj was discovered on 2004 July 31.76 in NGC 2403 \citep{2004IAUC.8377....1N} with
early observations in \cite{Vinko2006},  \citet{2008PZ.....28....8T} and \citealt{2009RAA.....9..783Z} with 
extensions to $\sim 1000$~days in \cite{Vinko2009}. 
\citet{2003PASP..115....1V} and \citet{2005MNRAS.360..288M} identify two progenitor candidates, and \citealt{2004ApJ...615L.113M} and 
\citealt{Vinko2009} estimate a mass of $\sim 15M_\odot$ based on the age of the associated star cluster, Sandage 96.
  There is evidence of CSM emission from early optical spectroscopy, late-time X-ray, and radio observations (\citealt{2007ApJ...662.1136C}, \citealt{2012ApJ...761..100C}, \citealt{2018ApJ...863..163N}). 
 Mid-IR Spitzer observations, late time optical photometry, and  HST data from days $\sim 106-1393$ after explosion provide evidence for dust formation in the ejecta (\citealt{2011A&A...527A..61S}, \citealt{2011ApJ...732..109M}).
 There are 52 LBT epochs for NGC~2403.

\subsection*{SN 2005cs}

SN 2005cs is a Type IIP SN in NGC 5194 (M 51) that was discovered on 2005 July 28.9 \citep{2005IAUC.8553....1K}. 
There are general observations of SN~2005cs in  \citet{2006MNRAS.370.1752P}, \citealt{2007ApJ...659.1488B} and \citet{2009MNRAS.394.2266P}.
Based on archival HST and Gemini images, the progenitor was identified as a K3 RSG, with an approximate mass of $(8\pm2)M_\odot$ \citep{2005MNRAS.364L..33M, 2006ApJ...641.1060L}. Early UV, optical, and a non-detection in the X-rays  provide an upper limit of $\dot{M} \ltorder 10^{-5}M_\odot$/year on the pre-SN mass loss rate (\citealt{2007ApJ...659.1488B}). There are 31 LBT epochs for NGC~5194.

\subsection*{SN 2009hd}

SN 2009hd occurred in NGC 3627 (M66) on 2009 July 2.69 \citep{2009CBET.1867....1M} and was classified as a Type IIL \citep{2011ApJ...742....6E}. This SN is located in a region with very high host extinction (see Table~\ref{tab:SN_INFO}). \citet{2011ApJ...742....6E} reports an upper limit on the luminosity of the progenitor of $L_{bol}\lesssim10^{5.04}L_{\odot}$ which we can use as an upper limit on the contributions of the progenitor to the
reference image.  The image subtraction residuals near the SN are unusually large, so while we have 27 LBT epochs, four of which were taken pre-outburst, the quantitative results are not very useful, particularly when combined with the
high extinction.

\subsection*{SN~2011dh}

SN~2011dh, was discovered in NGC 5914 (M51) on 2011 May 31.89 \citep{2011CBET.2736....1G, 2011ATel.3398....1S} and classified as a Type IIb SN. There are general studies of SN~2011dh in
\citet{2011ApJ...742L..18A}, \citet{2013MNRAS.433....2S}, \cite{2014A&A...562A..17E}, and \cite{2015A&A...580A.142E}.
The progenitor was yellow super-giant (YSG) (\citealt{2011ApJ...739L..37M}, \citealt{2011ApJ...741L..28V})
with properties discussed in \S3.1 and it appears to have been slightly variable (\citealt{2012ApJ...747...23S}).
\citealt{2014ApJ...793L..22F} argue that blue star detected in deep Near-UV HST images is a binary companion to the 
progenitor, but this is disputed by \cite{Maund2015}.
X-ray (\citealt{2014ApJ...785...95M}) and radio (\citealt{2012ApJ...752...78S}) observations found evidence
for CSM interactions, with a mass loss rate of order $\dot{M} \sim 3 \times 10^{-6} M_\odot$/year for $v_w=20$~km/s.
VLBI observations monitored the expansion of the remnant to a linear radius of $(7.4\pm0.3)\times10^{16}$cm implying an average expansion velocity of $\sim19000$~km/s \citep{2016MNRAS.455..511D}.
We have five observations before the SN and 26 afterwards, for a total of 31 LBT epochs.

\subsection*{SN 2012fh}

This system was discovered on 2012 October 18.856 \citep{2012CBET.3263....1N} in NGC 3344, and was classified as a Type Ic by \citet{2012CBET.3263....2T} and \citet{2012CBET.3263....3T}. As in \citet{2017MNRAS.472.3115J}, we will use a more generic Type Ibc classification because the spectra were collected long after peak. Photometry for 2102fh is available
in \cite{Shivvers2019} and \cite{Zheng2022}.  \citet{2017MNRAS.472.3115J} place limits on the luminosity of the progenitor,
finding that it likely had to be a lower mass star stripped by binary interactions. We have 10 LBT epochs  prior to the SN and 16 afterwards for a total of 26 epochs
This SN is the only candidate for which a binary companion analysis is carried out.

\subsection*{SN 2013am}

This Type II-P SN was discovered on 2013 Mar. 21.638 in NGC 3623 by \citet{2013CBET.3440....1N} and classified as a young Type II SN \citep{2013ATel.4909....1B}. There are a general studies of it in  \citet{2014ApJ...797....5Z} and \citet{2018MNRAS.475.1937T},
additional light curve data in \citet{2016AJ....151...33G} and \citet{2019MNRAS.490.2799D} and additional spectra
in \citet{2017MNRAS.467..369S}.  The spectrum in  \citet{2017MNRAS.467..369S} at $257$~days appears to be developing
the box-shaped line profiles typical of CSM interactions.  We have 8 pre-SN epochs of LBT data, one just 5 days before the discovery, and 29 epochs in total.

\subsection*{SN 2013ej}

This SN was discovered on 2013 July 24.83 by \citet{2013ATel.5466....1L} in NGC 628 (M74) and classified as a young Type II \citep{2013CBET.3606....1K}.  It is generally discussed in \citet{2014MNRAS.438L.101V}, \citet{2015ApJ...807...59H},
\citet{2015ApJ...807...59H}, \citealt{2016ApJ...822....6D} and \citet{2016MNRAS.461.2003Y} where the latter argues that it should be classified as a
Type~IIL rather than a Type~IIP.   \citet{2014MNRAS.439L..56F} argue that a red source in archival HST images might be the progenitor.  Early optical observations (\citealt{2015ApJ...806..160B}), X-ray observations (\citealt{2016ApJ...817...22C}) and spectropolarimetry (\citealt{2017ApJ...834..118M}) all support the presence
of CSM interactions.  We have 13 pre-SN epochs of LBT data and 35 in total.

\subsection*{ASASSN-16fq/SN~2016cok}

ASAS-SN \citep{2014ApJ...788...48S, 2017PASP..129j4502K} discovered the Type II-P SN ASASSN-16fq (SN 2016cok) in NGC 3627 (M66) on 2016 May 28.30  \citet{2016ATel.9091....1B}.  There is photometry of this SN in \citet{2019MNRAS.490.2799D}. \citet{2017MNRAS.467.3347K} estimate that the progenitor  was a $8 - 12M_{\odot}$ RSG with a luminosity of  $10^{4.5}$ - $10^{4.9}$ and that it had no significant pre-SN variability.
The SN was not detected in 23~ksec of Swift observations, but there is no discussion of the implications (\citealt{2016ATel.9091....1B}). We have 16 LBT epochs of pre-SN LBT data, and 26 in total.

\subsection*{SN 2017eaw}

SN 2017eaw is a Type II-P discovered on 2017 May 14.24 \citep{2017CBET.4390....1W, 2017ATel10372....1D} in NGC 6946. 
There are general studies of the SN and its progenitor in \citet{2018MNRAS.481.2536K}, \citet{2019ApJ...875..136V}, and 
\citet{2019ApJ...876...19S}.  There is additional photometry in \cite{2019MNRAS.487..832B} and evidence for dust
formation after the explosion (\citealt{2018ApJ...864L..20R}, \citealt{2019ApJ...873..127T}).  Late time spectra
show clear evidence of CSM interactions (\citealt{2020ApJ...900...11W}).
The progenitor luminosity is roughly $10^{4.9}L_\odot$ with a dusty wind or shell
(\citealt{2018MNRAS.481.2536K}). The 40 pre-SN LBT observations have no evidence for optical variability from the progenitor  over its last roughly decade (\citealt{2018MNRAS.480.1696J}), which essentially rules out suggestions
for late pre-SN outbursts (\citealt{2018MNRAS.481.2536K}, \citealt{2019MNRAS.485.1990R}).  
We have 40 pre-SN LBT epochs, and a total of 54. 


\bsp	
\label{lastpage}
\end{document}